\documentclass[11pt,letterpaper]{article}

\usepackage{latexsym}
\usepackage{epsfig}                     
\usepackage[usenames,dvipsnames]{color}
\usepackage{graphicx}
\graphicspath{{./images/}}
\usepackage{amssymb, amsmath, amsthm}   
\usepackage{enumerate}                  
\usepackage{mathrsfs}                   
\usepackage{multirow}
\usepackage{algorithm}
\usepackage{algorithmic}
\usepackage{longtable}
\usepackage{setspace}
\usepackage{placeins}
\usepackage{ wasysym }
\usepackage{geometry}                                                                          

\usepackage[T1]{fontenc}
\usepackage{lmodern}

\usepackage[backend=bibtex,citestyle=numeric-comp,hyperref=true,firstinits=true,terseinits=true,sorting=none,doi=false,url=false,maxbibnames=10]{biblatex}
\DeclareFieldFormat{postnote}{#1} 

\DeclareFieldFormat[article, inbook, incollection, inproceedings, patent, thesis, unpublished]{title}{#1}
\DeclareFieldFormat[article, inbook, incollection, inproceedings, patent, thesis, unpublished]{journaltitle}{\mkbibemph{#1}\nopunct}
\DeclareFieldFormat[article, inbook, incollection, inproceedings, patent, thesis, unpublished]{volume}{\mkbibparens{#1}\addcolon} 
\DeclareFieldFormat[article, inbook, incollection, inproceedings, patent, thesis, unpublished]{year}{\mkbibparens{#1}\nopunct} 

\DeclareFieldFormat[article, incollection, patent, thesis, unpublished]{pages}{{\nopp#1}}

\DeclareFieldFormat{sentencecase}{\MakeSentenceCase{#1}}
\renewbibmacro*{title}{%
  \ifthenelse{\iffieldundef{title}\AND\iffieldundef{subtitle}}
    {}
    {\ifthenelse{\ifentrytype{article}\OR\ifentrytype{inbook}%
      \OR\ifentrytype{incollection}\OR\ifentrytype{inproceedings}%
      \OR\ifentrytype{inreference}}
      {\printtext[title]{%
        \printfield[sentencecase]{title}%
        \setunit{\subtitlepunct}%
        \printfield[sentencecase]{subtitle}}}%
      {\printtext[title]{%
        \printfield[titlecase]{title}%
        \setunit{\subtitlepunct}%
        \printfield[titlecase]{subtitle}}}%
     \newunit}%
  \printfield{titleaddon}}

\DefineBibliographyStrings{english}{
urlseen = {Accessed:},
editor = {(Ed)},
editors = {(Eds)},
byeditor = {(Ed.)}}

\DeclareNameAlias{default}{last-first}

\defbibheading{siref}[\refname]{\section{#1}}

\renewbibmacro{in:}{}

\renewbibmacro{publisher+location+date}{
  \iflistundef{publisher}
    {}
    {\printlist{publisher}%
       {\addcomma\space}%
      \iflistundef{location}
        {}
        {\printlist{location}}%
    }
}

\DeclareBibliographyDriver{article}{%
\usebibmacro{bibindex}%
\usebibmacro{begentry}%
\usebibmacro{author/translator+others}%
\newunit\newblock
\printfield{year}%
\setunit{\labelnamepunct}\newblock
\usebibmacro{title}%
\newunit
\printlist{language}%
\newunit\newblock
\usebibmacro{byauthor}%
\newunit\newblock
\usebibmacro{bytranslator+others}%
\newunit\newblock
\printfield{version}%
\newunit\newblock
\usebibmacro{journal}%
\newunit\newblock
\printfield{volume}%
\newunit\newblock
\usebibmacro{byeditor+others}%
\newunit\newblock
\usebibmacro{note+pages}%
\newunit\newblock
\iftoggle{bbx:isbn}
{\printfield{issn}}
{}%
\newunit\newblock
\usebibmacro{doi+eprint+url}%
\newunit\newblock
\usebibmacro{addendum+pubstate}%
\newunit\newblock
\usebibmacro{pageref}%
\usebibmacro{finentry}}

\DeclareBibliographyDriver{inproceedings}{%
\usebibmacro{bibindex}%
\usebibmacro{begentry}%
\usebibmacro{author/translator+others}%
\newunit\newblock
\printfield{year}%
\setunit{\labelnamepunct}\newblock
\usebibmacro{title}%
\newunit
\printlist{language}%
\newunit\newblock
\usebibmacro{byauthor}%
\newunit\newblock
\usebibmacro{bytranslator+others}%
\newunit\newblock
\printfield{version}%
\newunit\newblock
\usebibmacro{booktitle}%
\newunit\newblock
\printfield{volume}%
\newunit\newblock
\usebibmacro{byeditor+others}%
\newunit\newblock
\usebibmacro{publisher+location+date}%
\newunit\newblock
\usebibmacro{note+pages}%
\newunit\newblock
\usebibmacro{pageref}%
\usebibmacro{finentry}}

\DeclareBibliographyDriver{book}{%
\usebibmacro{bibindex}%
\usebibmacro{begentry}%
\usebibmacro{author/translator+others}%
\newunit\newblock
\printfield{year}%
\setunit{\labelnamepunct}\newblock
\usebibmacro{title}%
\newunit
\printlist{language}%
\newunit\newblock
\usebibmacro{byauthor}%
\newunit\newblock
\usebibmacro{bytranslator+others}%
\newunit\newblock
\usebibmacro{booktitle}%
\newunit\newblock
\printfield{volume}%
\newunit\newblock
\usebibmacro{publisher+location+date}%
\newunit\newblock
\usebibmacro{note+pages}%
\newunit\newblock
\usebibmacro{pageref}%
\usebibmacro{finentry}}

\definecolor{citescol}{RGB}{0,17,202}
\definecolor{urlscol}{RGB}{0,0,0}
\definecolor{linkscol}{RGB}{194,0,10}
\definecolor{headcol}{RGB}{0,9,160}
\definecolor{subheadcol}{RGB}{0,79,160}

\usepackage{hyperref}
\usepackage[all]{hypcap}
\hypersetup{colorlinks=true,linkcolor=linkscol,citecolor=citescol,urlcolor=urlscol}

\definecolor{purpcol}{RGB}{195,9,235}

\usepackage{bm}

\theoremstyle{definition}

\theoremstyle{definition}

\theoremstyle{definition}

\newcommand{\citep}{\cite}
\newcommand{\citet}{\cite}

\definecolor{lg}{gray}{0.75}
\def\gcirc{{%
    \setbox0\hbox{$\fullmoon$}%
    \rlap{\hbox to \wd0{\hss{$\textcolor{lg}{\newmoon}$}\hss}}\box0
}}
\def\gtri{{%
    \setbox0\hbox{\large$\vartriangle$}%
    \rlap{\hbox to \wd0{\hss{\large$\textcolor{lg}{\blacktriangle}$}\hss}}\box0
}}

\begin{document}

\begin{flushright}
\href{http://arxiv.org/abs/1310.2968}{arXiv:1310.2968}
\end{flushright}
\vskip1cm

\vskip1cm
\begin{center}

\noindent{\Large \bf The Fossilized Birth-Death Process:\\ 
\vskip0.3cm
A Coherent Model of Fossil Calibration \\
\vskip0.3cm
 for Divergence Time Estimation}
\vskip1cm

\noindent{\normalsize \sc Tracy A.\ Heath$^{1,2}$, John P.\ Huelsenbeck$^{1,3}$, and Tanja Stadler$^{4,*}$}
\bigskip

\noindent{\small \it 
$^1$Department of Integrative Biology, University of California, Berkeley, CA 94720 USA;\\
$^2$Department of Ecology and Evolutionary Biology, University of Kansas, Lawrence, KS 66045 USA;\\
$^3$Department of Biological Sciences, Faculty of Science, King Abdulaziz University, Jeddah 21589, Saudi Arabia;\\
$^4$Department of Environmental Systems Science, Eidgen\"{o}ssische Technische Hochschule Z\"{u}rich, 8092 Z\"{u}rich, Switzerland.\\
}
\end{center}
\medskip
\noindent{$^*$\bf{Corresponding author:}} T.\ Stadler, Institut f.\ Integrative Biologie,
CHN H 75.1,
Universit\"{a}tstrasse 16, 8092 Z\"{u}rich, Switzerland; E-mail: \href{mailto:tanja.stadler@env.ethz.ch}{tanja.stadler@env.ethz.ch}.\\

\newpage

\begin{abstract}
Time-calibrated species phylogenies are critical for addressing a wide range of questions in evolutionary biology, such as those that elucidate historical biogeography or uncover patterns of coevolution and diversification. 
Because molecular sequence data are not informative on absolute time, external data---most commonly fossil age estimates---are required to calibrate estimates of species divergence dates. 
For Bayesian divergence-time methods, the common practice for calibration using fossil information involves placing arbitrarily chosen 
parametric distributions on internal nodes, often disregarding most of the information in the fossil record. 
We introduce the `fossilized birth-death' (FBD) process --- a model for calibrating divergence-time estimates in a Bayesian framework, explicitly acknowledging that extant species and fossils are part of the same macroevolutionary process. Under this model, absolute node age estimates are calibrated by a single diversification model and arbitrary calibration densities are not necessary. Moreover, the FBD model allows for inclusion of all available fossils. We performed analyses of simulated data and show that node-age estimation under the FBD model results in robust and accurate estimates of species divergence times with realistic measures of statistical uncertainty, overcoming major limitations of standard divergence time estimation methods. 
We then used this model to estimate the speciation times for a dataset composed of all living bears, indicating that the genus \textit{Ursus} diversified in the late Miocene to mid Pliocene. 
\end{abstract}

\noindent (Keywords: Bayesian divergence time estimation, relaxed-clock, fossil calibration, MCMC, birth-death process)

\newpage



\section{Introduction}
A phylogenetic analysis of species has two goals: to
infer the evolutionary relationships and the amount of divergence among species. 
Preferably, divergence is estimated in units proportional to time, thus revealing the times at which speciation events occurred. 
Once orthologous DNA sequences from the species have been aligned,
both goals can be accomplished by assuming that nucleotide substitutions
occur at the same rate in all lineages 
(the `molecular-clock' assumption \cite{zuckerkandl62}) and that the time of at least one speciation event on the tree is known, {\it i.e.}, one
speciation event acts to `calibrate' the rate of substitution.

The goal of reconstructing rooted, time-calibrated phylogenies is complicated by
substitution rates changing over the tree
and by the difficulty of determining
the date of any speciation event.
Substitution-rate variation among lineages is pervasive and has been accommodated in a number of ways. 
The most widely
used method to account for rate heterogeneity is to assign
an independent parameter to each branch of the tree.
Branch lengths, then, are
the product of substitution
rate and time, and usually measured in units of expected number of
substitutions per site. 
This solution allows estimation of the tree topology --- which is informative about inter-species relationships --- but
does not attempt to estimate the rate and time separately.
Thus, under this `unconstrained' parameterization, molecular-sequence data allow inference of phylogenetic relationships and genetic distances among species, but the timing of speciation events is confounded in the branch-length parameter \cite{rannala2003,thorne05,dosReis2013}.
Under a `relaxed-clock' model, substitution rates change over the tree in a constrained manner, thus separating the rate and time parameters associated with each branch and allowing inference of lineage divergence times.
A considerable amount of effort has been directed at modeling lineage-specific substitution rate variation, with many different
relaxed-clock models described in the literature \cite{hasegawa89,kishino90,thorne98,huelsenbeck00a,yoder00,kishino01, thorne02,arisbrosou02,yang03,drummond06,lepage06, lepage07,rannala07,drummond10,heath11}. 
When such models are coupled with a model on the distribution of speciation events over time (\textit{e.g.}, the Yule model \cite{yule24} or birth-death process \cite{kendall48}), molecular-sequence data can then inform the \textit{relative} rates and node ages in a phylogenetic analysis. 

Estimates of branch lengths in units of \textit{absolute} time (\textit{e.g.}, millions of years) are required for studies investigating comparative or biogeographical questions (\textit{e.g.}, \cite{antonelli2011,cardinal2013}).
However, because commonly used diversification priors are imprecise on node ages, external information is required to infer the absolute timing of speciation events.
Typically, a rooted time tree is calibrated by constraining the ages of a set of internal nodes. 
Age constraints may be derived from several sources, but the most common and reliable source of calibration information is the fossil record \cite{marshall08,kodandaramaiah10}. 
Despite the prevalence of these data in divergence-time analyses, the problem of properly calibrating a phylogenetic tree has received less consideration
than the problem of accommodating variation in substitution rate.
Moreover, various factors can lead to substantial errors in parameter estimates \cite{graur04,benton03,gandolfo08,ho09,lloyd12,heath12}.
When estimating node ages, calibration nodes are identified, representing the most-recent common ancestor (MRCA) of a fossil and set of extant species. Based on the fossil, the calibration node's age is estimated.
Thus, fossil data can typically only provide valid minimum-age constraints on these nodes \cite{benton03,marshall08}, and erroneous conclusions can result if the calibrated speciation event is not properly specified \cite{graur04}.

Bayesian inference methods are well adapted to accommodating uncertainty in calibration times 
by assuming that the age of the calibrated node is a random variable
drawn from some parametric probability distribution 
\citep{kishino01,drummond06,yang06,benton07,ho07,ho09,heath12,heled12}.
Although this Bayesian approach properly propagates uncertainty in the calibration times through the analysis (reflected in the credible intervals on uncalibrated node ages), 
two problems remain unresolved. 
First, these approaches, as they are commonly applied, induce a probability distribution on the age of each calibrated node that comes from both the node-specific calibration prior and the tree-wide prior on node ages, leading to an incoherence in the model of branching times on the tree \cite{heled12,warnock12}.
Typically, a birth-death process of cladogenesis is considered as the generating model for the tree and speciation times
\cite{yule24,kendall48,rannala96,yang97b,gernhard08,stadler09}, serving as the tree-wide prior distribution on
branch times throughout the tree in a Bayesian analysis. 
Those speciation events acting as  calibrations
are then considered to be drawn from an additional, unrelated probability distribution intended to
model uncertainty in the calibration time. 
Importantly, this problem is avoided by partitioning the nodes and applying a birth-death process to uncalibrated nodes conditioned on the calibrated nodes \cite{yang06}, although many divergence-time methods do not use this approach.
Nevertheless, a single model 
that acts as a prior on the speciation times for both calibrated and uncalibrated nodes is a better representation of the lineage-diversification process and preferable as a prior on branching times when using fossil data. 
Second, the probability distributions used to model uncertainty in calibration times are
poorly motivated. 
The standard practice in Bayesian divergence-time methods is to model uncertainty in calibrated node ages using simple probability distributions, such
as the uniform, log-normal, gamma, or exponential distributions \cite{ho09}. 
When offset by a minimum age, these `calibration densities' \cite{heled12} simply seek to characterize the age of the node with respect to its descendant fossil. 
However, the selection and parameterization of calibration priors are rarely informed by any biological process or knowledge of the fossil record (except see \cite{lee09,wilkinson11,nowak2013}). 
A probability model that acts as a fossil-calibration prior should have parameters
relevant to the preservation history of the group, such as the rate at which fossils occur in the rock record, a task that is likely to be difficult for most groups without an abundant fossil record \cite{wilkinson11,wagner2013}. 
Consequently, most biologists are faced with the challenge of choosing and parameterizing calibration densities without an explicit way of describing their prior knowledge about the calibration time. 
Thus, calibration priors are often specified based on arbitrary criteria or \textit{ad hoc} validation methods \cite{dornburg11}; and ultimately, this may lead to arbitrary or \textit{ad hoc} estimates of divergence times. 

We propose a novel way of calibrating phylogenies with fossils. 
Since molecular sequences from extant species and fossils are different observations of the same speciation and extinction process, we use an explicit speciation-extinction-fossilization model to describe the distribution of speciation times and recovered fossils. 
This model --- the fossilized birth-death (FBD) process  --- acts as a prior for divergence-time dating.
The parameters of the model --- the 
speciation rate, extinction rate, fossil recovery rate, and proportion of sampled extant species ---  interact to inform the amount of uncertainty
for every speciation event on the tree. 
These four parameters are the only quantities requiring prior assumptions, compared with assuming separate calibration densities for each fossil. 

\section{Results and Discussion}

\subsection{A Unified Model for Fossil and Extant Species Data}

\begin{figure}[tbh!]
\hfil\includegraphics[scale=0.5]{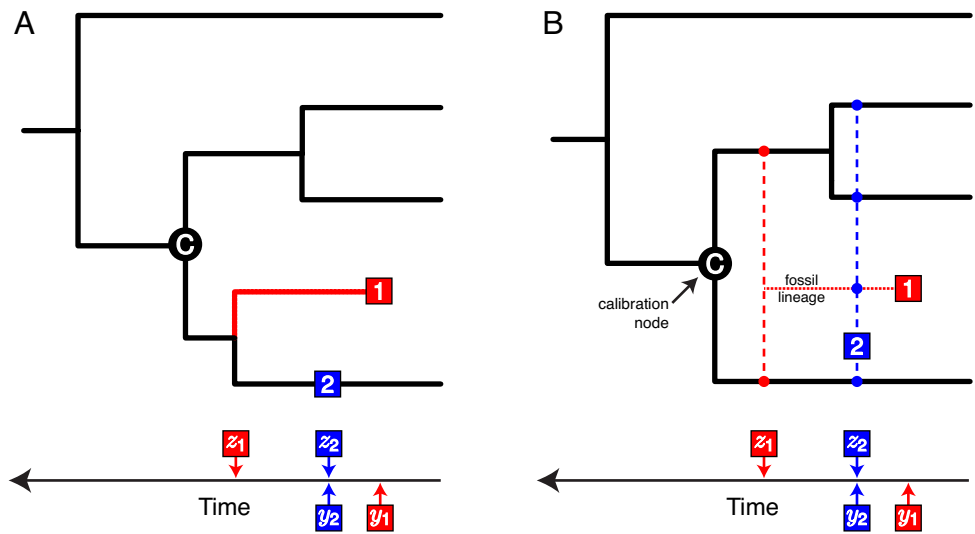}\hfil
\caption{(A) The phylogeny of four extant species and two sampled fossils, where the true phylogenetic relationships are known. This tree is the serially sampled birth-death (SSBD) tree described in Stadler \cite{stadler10}. 
Each fossil has an observed age, $y$, and an attachment time, $z$, which is the point at which the fossil links to the tree.
Fossil 1 (red), the youngest fossil specimen, is descended from node C via speciation at time $z_1$ and occurs on an extinct lineage, such that $y_1 < z_1$. 
Fossil 2 (blue) lies directly on a lineage in the extant tree, therefore $y_2 = z_2$ and fossil 2 is the ancestor of a sampled, present-day taxon.
(B) The fossilized birth-death (FBD) tree where the precise phylogenetic relationships of the fossils are ignored and the two fossil specimens are used to calibrate the extant tree. 
Both fossils calibrate a single internal node, C, 
because there is prior knowledge that fossils 1 and 2 are descendants of the calibration node. All other nodes are uncalibrated.
Because fossil 1 (red) attaches to the extant tree via speciation, where $y_1 < z_1$, the unobserved speciation event is assumed to occur at any lineage that is descended from C and that intersects with $z_1$ (small, red circles). 
The attachment time of fossil 2 (blue) is equal to the age of the fossil, thus fossil 2 represents a direct ancestor of any lineage descended from node C that intersects with time $z_2$ (small, blue circles), including any ``ghost'' lineages leading to other fossil taxa (\textit{e.g.}, the dotted, red line leading to fossil 1). 
}\label{fbdTree}
\end{figure}

The fossilized birth-death (FBD) model gives rise to time-calibrated phylogenies of extant species, together with occurrence times and attachment ages of sampled fossils (Fig.~\ref{fbdTree}). 
This model is derived from the \textit{serially sampled birth-death} (SSBD) process \cite{stadler10,didier2012}, which defines a rooted phylogeny of $n$ extant tips and $m$ sampled fossils (Fig.~\ref{fbdTree}A). 
The process of diversification under the SSBD model starts with a single lineage at time $x_0$ (stem age) before the present. 
The model assumes a constant speciation rate $\lambda$, and a constant extinction rate $\mu$. Recovered fossils appear along lineages of the complete species tree according to a Poisson process with parameter $\psi$. 
Finally, each extant species is sampled with probability $\rho$.
This process gives rise to trees with extinct and extant tips and fossilization occurrences along the branches. 
Deleting all lineages without sampled extant or fossil descendants leads to the {\it sampled phylogeny} (Fig.~\ref{fbdTree}A). Deleting all lineages without sampled extant descendants leads to the {\it reconstructed phylogeny} (Fig.~\ref{fbdTree}A, black).
We denote the age of the $i^{th}$ internal node in the reconstructed phylogeny with $x_i$ (for $i\in 1,\ldots, n-1$). The age of fossil $f$ is denoted with $y_f$ and the attachment time of $f$ to the reconstructed phylogeny is $z_f$.

The SSBD model requires that the phylogenetic relationships of both extant and fossil taxa are explicitly represented (Fig.~\ref{fbdTree}A).
Accordingly, this model is appropriate for describing the distribution of speciation times and tree topologies when used in `tip-dating' approaches, where sequence data --- either molecular or discrete morphological characters --- are available for both extant \textit{and} extinct taxa \cite{drummond2003,Shapiro2004,pyron11,ronquist11,Stadler012,StadlerYang2013}. 
However, suitable data matrices of discrete morphological characters for both living and fossil taxa are unavailable for many groups in the tree of life. 
Instead, biologists may only have access to the times of fossil occurrences and their taxonomic identification based on a few diagnostic characters. 
In these cases, fossils are still useful for calibrating a molecular phylogeny of extant species provided that calibration nodes are correctly identified. 
Though, conventional calibration-density approaches for Bayesian divergence time estimation require that prior densities are parameterized for each calibrated node \cite{ho09}, a task that presents a challenge for most biologists performing these analyses.

The FBD model overcomes several of the major limitations associated with standard node-calibration for phylogenetic datasets unsuitable for tip dating.
This is achieved by adapting the SSBD model to account for ambiguity in the precise phylogenetic placement of fossils while still considering them part of a unified macroevolutionary process. 
Moreover, like tip-dating and the SSBD model, divergence-time estimation under the FBD model allows for inclusion of \textit{all} reliable fossil taxa available for the group of interest and eliminates the need for \textit{ad hoc} calibration prior densities without requiring combined character data for both modern and fossil species. 

When the FBD model is applied as a prior on divergence times in a Bayesian framework, we do not require nor infer information on the exact phylogenetic relationships of fossils. Instead, we infer {\it FBD trees}, defined as dated, extant-species phylogenies together with the age and time of attachment for each fossil in a sampled tree (Fig.~\ref{fbdTree}B). 
For any fossil, if $y_f < z_f$, then fossil $f$ attaches to the sampled tree by way of speciation and induces an unobserved lineage. 
Alternatively, Foote \cite{foote1996} described models to assess the probability of ancestor-descendant pairs in the fossil record, and under the FBD model, this pattern is possible when $y_f = z_f$, such that fossil $f$ lies directly on a branch and is an ancestor of sampled extant or fossil taxa (Fig.~\ref{fbdTree}B). 
The precise topology of the sampled phylogeny is ignored when calculating the probability of the FBD tree by summing over the probabilities of all possible sampled phylogenies that can induce a given FBD tree (see Methods \ref{FBDprobSec}; Eq.~\ref{EqnTreeLik}).
Importantly, since we do not know if a given fossil is the direct ancestor of a lineage in the FBD tree or if it lies on an unobserved lineage, we can average over all possible FBD tree realizations using numerical methods.
In order to use the FBD model as a prior for divergence-time dating,  we calculate the probability density for obtaining a particular sampled phylogeny under the SSBD model, conditioning on the root (crown) age of the tree, $x_1$ (Eq.~\ref{EqnTreeLik}).

The FBD tree probability is central to our Bayesian divergence-time estimation method implemented in the program DPPDiv (\href{https://github.com/trayc7/FDPPDIV}{https://github.com/trayc7/FDPPDIV}) \cite{heath11,heath12,Darriba2013}. 
This approach estimates FBD trees -- specifically divergence times -- from molecular sequence data for extant species together with fossil occurrence times. 
The user provides only the sequence data, extant tree topology, fossil ages, and calibration nodes for each fossil. 
We use Markov chain Monte Carlo (MCMC) to approximate the posterior distribution of FBD trees conditional on the extant species tree topology and nodal assignments for each fossil specimen (Methods \ref{MCMCMethodSec}). 
Note that, like calibration-density methods, divergence-time estimation under the FBD model requires that the fossil is correctly assigned to a node in the extant tree that is truly older than the fossil age.
However, in contrast to calibration-density approaches that require the user to choose and parameterize a prior density for each calibrated node, the only input assumptions required when applying the FBD model are prior information on the FBD model parameters ($\lambda, \mu, \psi, \rho$), 
parameters of the GTR substitution model with gamma-distributed rate heterogeneity, and parameters of the model of lineage-specific rate variation (\textit{e.g.}, relaxed-clock model). 
We used both simulated and empirical data to evaluate the accuracy, precision, and robustness of divergence-time estimation under the FBD model.
Our simulation results show that integrating fossil information into the diversification model yields accurate inferences of absolute node ages.
Furthermore, the FBD model provides coherent measures of statistical uncertainty that lead to more straightforward interpretation compared with standard practices for node calibration.

\subsection{Analyses of Simulated Data}

\subsubsection{Accuracy and Precision of Node-Age Estimates}
We generated tree topologies and sets of fossils under a forward-time simulation of the constant-rate SSBD model. 
Our results are focused on a set of simulations with the following conditions: $r =0.5$, $d = 
0.01$, and $\psi = 0.1$; where $r$ is the turnover rate, such that $r=\mu/\lambda$; the diversification rate is $d=\lambda-\mu$; and $\psi$ is the Poisson rate of fossilization events (see Methods \ref{MethSimDetails} for simulation details). 
Each of our 100 simulation replicates comprised a tree topology for 25 extant species with corresponding sequence data (GTR+$\Gamma$, strict molecular clock), and a complete set of fossil ages, each associated with the true MRCA in the extant tree. 
We then subsampled each set of complete fossils, so that only a percentage, $\omega$, of the fossils were present. 
We produced four different fossil subsets under different values of $\omega$: 5\%, 10\%, 25\%, and 50\% (henceforth we use the notation $\omega_P$ to indicate $\omega = P\%$).
Additionally, for sets of $\omega _{10}$ fossils (across 100 replicates: median = 17 fossils), we created a set of `calibration fossils' ($\omega_{cal10}$), where, for each calibrated node, only the oldest fossils were retained (median = 9 fossils).
We created the $\omega_{cal10}$ fossils to compare node calibration under the FBD model with commonly used calibration density approaches which do not consider all available fossils. 

We estimated absolute node ages for trees of 25 extant species under the FBD model using each set of fossils: $\omega_5$, $\omega_{10}$, $\omega_{25}$, $\omega_{50}$, and $\omega_{cal10}$ (Methods \ref{MethCalFBDAn}).
For comparison, we estimated node ages under three different calibration-density approaches using the $\omega_{cal10}$ set of fossils, each assuming an exponential calibration density (Methods \ref{MethCalDenAn}).
The \textsl{fixed-scaled} calibration density method scaled the expected value of each exponential prior density based on the age of the calibrating fossil. 
Another calibration density was created where the expected age of the calibrated node was equal to the \textit{true} age (\textsl{fixed-true}). 
The third calibration-density approach applied a  \textsl{hyperprior} to the rate parameters of exponential distributions, such that the nodes are calibrated by a mixture of exponentials \cite{heath12}.
The fixed-true calibration density is expected to perform well and represent an ideal prior density parameterization.
The hierarchical, hyperprior calibration density approach has been evaluated using simulated data and yields accurate estimates of node ages \cite{heath12}.
The fixed-scaled approach is intended to mimic arbitrary parameter specification, though, it may lead to overly informative calibration densities.
For each analysis, we estimated absolute node ages in the program DPPDiv \cite{heath11,heath12,Darriba2013}, conditioning on the true tree topology and assuming a GTR+$\Gamma$ model of sequence evolution and a strict molecular clock (true models).

\begin{figure}[tbh!]
\hfil\includegraphics[scale=0.6,angle=-90]{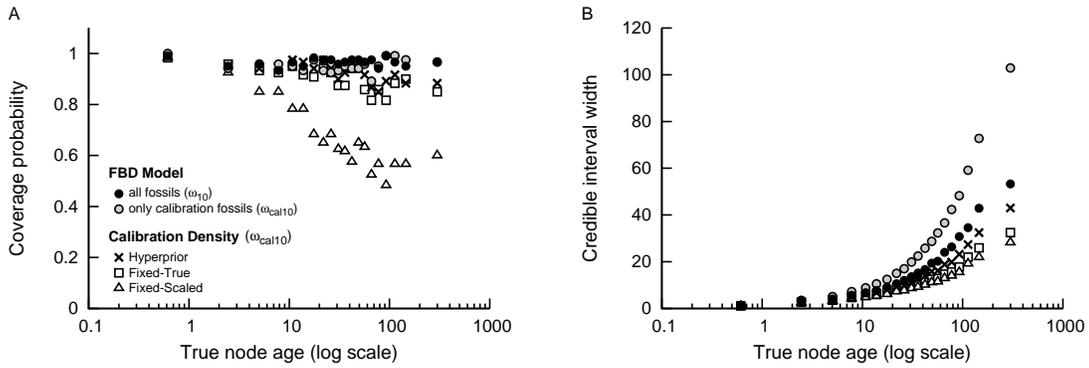}\hfil
\caption{The results for 100 replicate trees simulated under the FBD model with $\mu/\lambda = 0.5$, $\lambda - \mu = 0.01$. Node-age estimates are summarized for analyses under the FBD model using all available fossils sampled randomly (10\%; $\omega_{10}$) from the total number of simulated fossils ($\newmoon$). These results are compared to divergence time estimates on the set of calibration fossils ($\omega_{cal10}$) under the FBD model (\gcirc), with a hyperprior on calibration-density parameters ($\times$), with a fixed calibration density where the expected value is equal to the true node age ($\square$), and for a fixed calibration density scaled based on the age of the fossil ($\bigtriangleup$). Both the coverage probability and precision (95\% CI width) are shown as a function of the true node age (log scale), where the nodes were binned so that each bin contained 100 nodes and the statistics were computed within each bin. (A) The coverage probability is the proportion of nodes where the true value falls within the 95\% credible interval. (B) The average size of the 95\% credible intervals for each bin were computed to evaluate precision.}
\label{simFig1}
\end{figure}

We compared the node-age estimates under the FBD model to the three calibration-density analyses (fixed-scaled, fixed-true, and hyperprior).
Our results show that estimates of absolute node ages are more reliable under the FBD model relative to calibration density approaches.
Figure \ref{simFig1}A shows the coverage probabilities for node age estimates under the FBD model, using both the $\omega_{10}$ and $\omega_{cal10}$ sets of fossils; and for the calibration-density analyses using only the $\omega_{cal10}$ fossils.
The coverage probability (CP) is the proportion of nodes for which the true value falls within the 95\% credible interval (95\% CI).
When calculated across all nodes, the CP for ages estimated under the FBD model was 0.966 for the $\omega_{10}$ fossils and 0.953 for the $\omega_{cal10}$ fossils, indicating robust inference under the FBD process (Table \ref{summaryTableCompare}).
On average, the more reliable calibration-density approaches had high coverage using the $\omega_{cal10}$ fossils, with CP = 0.93 under the hyperprior calibrations, 
and CP = 0.904 when the expectation of the calibration density was equal to the true node age (fixed-true).
By contrast, the calibration densities parameterized based on the magnitudes of fossil ages (fixed-scaled) had relatively low CP across all nodes: CP = 0.68.
In Figure \ref{simFig1}A we show CP as a function of the true node age by creating bins of 100 nodes and computing the CP for each bin.
The node age estimates under the FBD model show consistently high coverage probabilities for both the $\omega_{10}$ and $\omega_{cal10}$ sets of fossils. 
In comparison, hyperprior and fixed-true calibration density analyses show slightly reduced CP for older nodes; and the fixed-scaled calibration densities have very low coverage probabilities as node age increases.

Figure \ref{simFig1}B illustrates the precision of divergence-time estimates by depicting the average 95\% CI width for increasing node age. 
These results show that estimates under the FBD model are less precise (\textit{i.e.}, larger 95\% CIs) relative to the calibration density methods. 
Therefore, inference under the FBD model yields conservative estimates of node ages, with high CPs combined with large 95\% CIs. 
In contrast, the precision of node-age estimates under calibration-density methods are primarily driven by the variance of calibration priors \cite{dosReis2013}. 
Importantly, however, the average widths of the 95\% CIs are smaller with the $\omega_{10}$ set of fossils compared to the reduced set of calibration fossils, $\omega_{cal10}$ (Fig.~\ref{simFig1}B).
Thus, as more fossils are added, precision under the FBD model increases.

\begin{figure}[tbh!]
\hfil\includegraphics[scale=0.6,angle=-90]{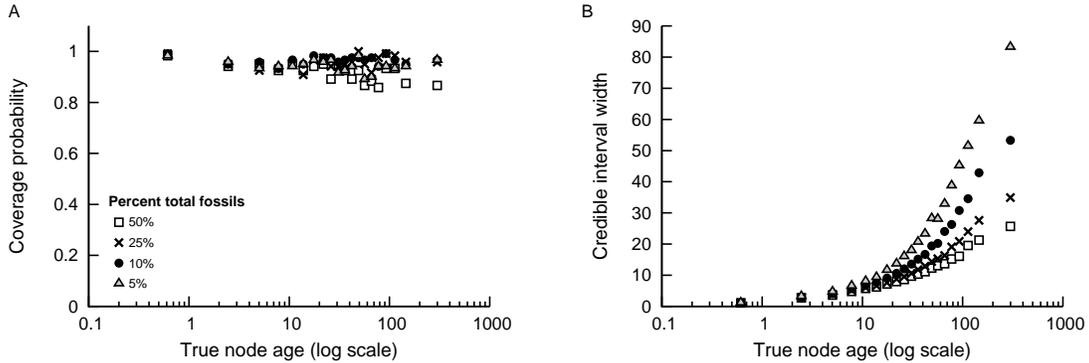}\hfil
\caption{The results for 100 replicate trees simulated under the FBD model with $\mu/\lambda = 0.5$, $\lambda - \mu = 0.01$. Node-age estimates are summarized for analyses under the FBD model using sets of fossils sampled from the total number of simulated fossils. Fossils were sampled at different percentages: 5\% (\gtri), 10\% ($\newmoon$), 25\% ($\times$), 50\% ($\square$). Both the coverage probability and precision (95\% CI width) are shown as a function of the true node age (log scale), where the nodes were binned so that each bin contained 100 nodes and the statistics were computed within each bin. (A) The coverage probability is the proportion of nodes where the true value falls within the 95\% credible interval. (B) The average size of the 95\% credible intervals for each bin were computed to evaluate precision.}
\label{simFig2}
\end{figure}

We further investigated the effect of different levels of fossil sampling under the FBD model. 
Figure \ref{simFig2} shows the results for node age estimation for different values of $\omega$: 5\%, 10\%, 25\%, 50\%.
Overall, we do not observe large changes to the coverage of estimates using different sets of fossils, with CPs remaining consistently high (Fig.~\ref{simFig2}A).
However, when we examine the precision of node-age estimation under the FBD model, we find that as the density of sampled fossils increases, the precision of divergence-time estimates also increases (Fig.~\ref{simFig2}B). 

\begin{figure}[tbh!]
\hfil\includegraphics[scale=0.5,angle=-90]{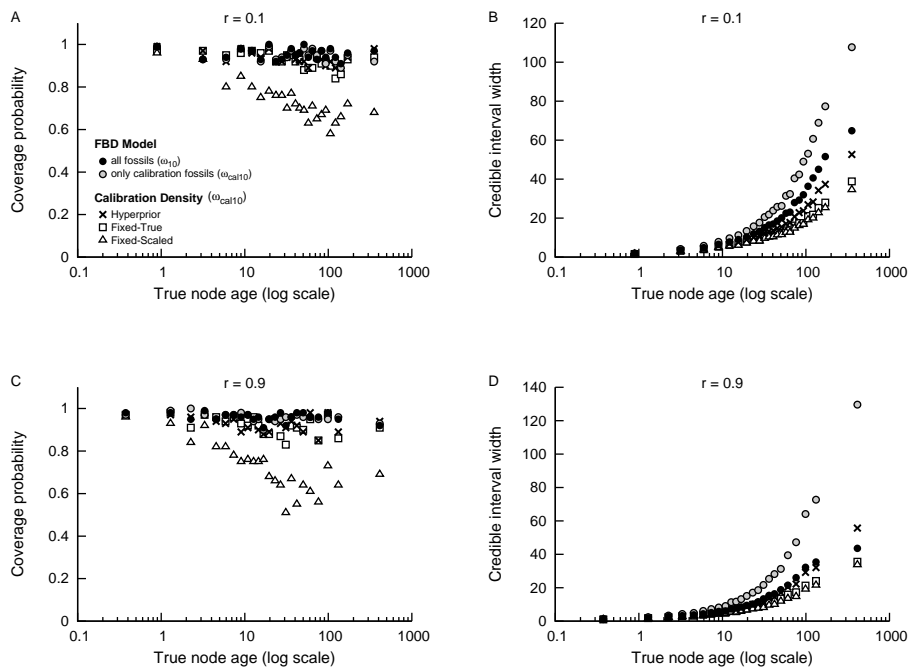}\hfil
\caption{The coverage probability (CP) and precision of node-age estimates for data simulated under different model parameters. 
The top row shows the results for $r = 0.1$, $\lambda - \mu = 0.0135$; and the second row is for trees with $r = 0.9$, $\lambda - \mu = 0.004$, where $r = \mu/\lambda$. 
Node-age estimates are summarized for analyses under the FBD model using all available fossils sampled randomly (10\%; $\omega_{10}$) from the total number of simulated fossils ($\newmoon$). 
These results are compared to divergence time estimates on the set of calibration fossils ($\omega_{cal10}$) under the FBD model (\gcirc), with a hyperprior on calibration-density parameters ($\times$), with a fixed calibration density where the expected value is equal to the true node age ($\square$), and for a fixed calibration density scaled based on the age of the fossil ($\bigtriangleup$). Both the CP and precision are shown as a function of the true node age (log scale), where the nodes were binned so that each bin contained 100 nodes and the statistics were computed within each bin. 
For (A) and (C), the coverage probability is the proportion of nodes where the true value falls within the 95\% credible interval. 
In (B) and (D), the average size of the 95\% credible intervals for each bin were computed to evaluate precision.}
\label{rndTurnoverCompareFig}
\end{figure}

\begin{figure}[tbh!]
\hfil\includegraphics[scale=0.5,angle=-90]{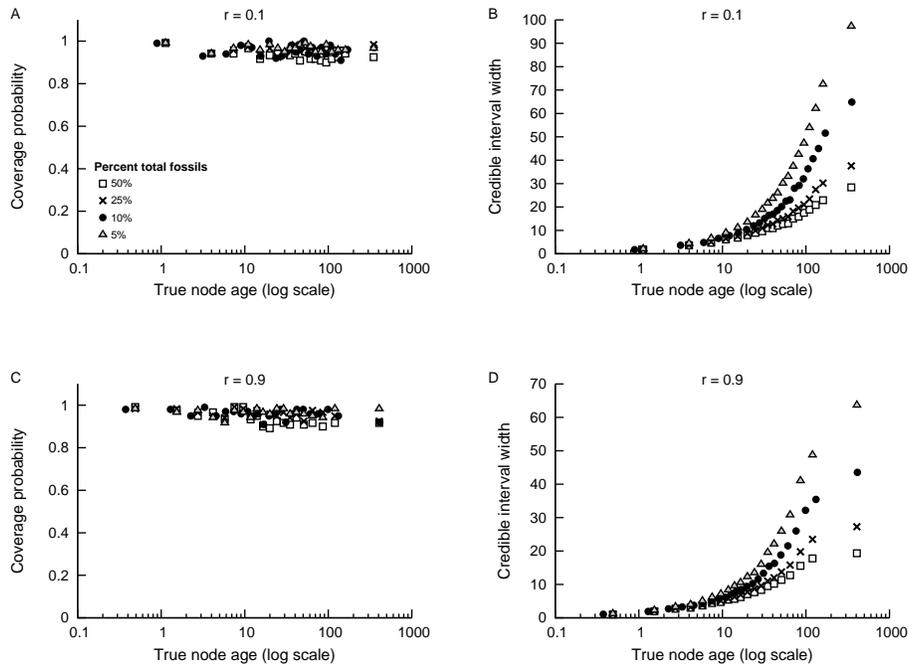}\hfil
\caption{The results for 100 replicate trees simulated under the FBD model with $r = 0.1$ and $r = 0.9$. Node-age estimates are summarized for analyses under the FBD model using sets of fossils sampled from the total number of simulated fossils. Fossils were sampled at different percentages: 5\% (\gtri), 10\% ($\newmoon$), 25\% ($\times$), 50\% ($\square$). Both the CP and precision are shown as a function of the true node age (log scale), where the nodes were binned so that each bin contained 100 nodes and the statistics were computed within each bin. 
For (A) and (C), the CP is the proportion of nodes where the true value falls within the 95\% credible interval. 
In (B) and (D), the average size of the 95\% credible intervals for each bin were computed to evaluate precision.}
\label{rndTurnoverFossSampFig}
\end{figure}

Our simulation results make a strong case for node-age calibration under the FBD model.
Furthermore, these patterns -- high coverage probabilities, with increased fossil sampling leading to increased precision -- are consistent under different simulation conditions. 
In particular, we focused on varying the turnover rate, $r = \mu / \lambda$, leaving the Poisson rate of fossilization at $\psi=0.1$, while changing the diversification rate, $d = \lambda - \mu$, to ensure that the expected root age, 
was approximately equal to 200 (varying $d$ can be seen as simply changing the time scale).
In Figures \ref{rndTurnoverCompareFig} and \ref{rndTurnoverFossSampFig}, the results for $r = 0.1$ and $r = 0.9$ exhibit patterns similar to those shown in Figures \ref{simFig1} and \ref{simFig2}, 
with high coverage across all estimates (Table \ref{summaryTableCompare}). 

\subsubsection{Robustness to Model Violation}

Much like calibration-density approaches, estimates of node ages under the FBD model may be sensitive to sampling of tips and fossils and parameterization of the model.
Therefore, we conducted additional simulations to examine conditions under which assumptions of the FBD model are violated. 

\begin{figure}[tbh!]
\hfil\includegraphics[scale=0.6,angle=0]{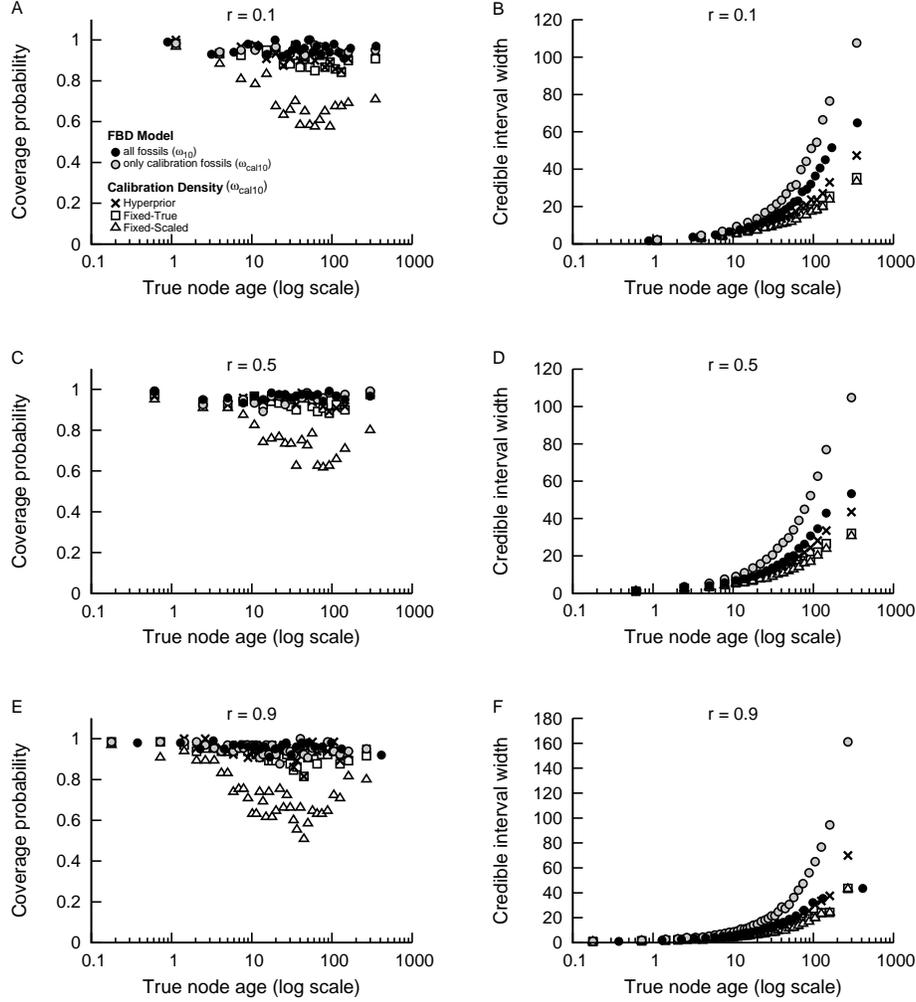}\hfil
\caption{The effect of preservation-biased fossil sampling on the CP and precision of node age estimates. 
Divergence times were estimated on all fossils sampled in the $\omega_{10}$ set of fossils under the FBD model ($\newmoon$).
Results using the set of calibration fossils ($\omega_{cal10}$) are shown for estimates under the FBD model (\gcirc) and calibration-density methods using a hyperprior on exponential-rate parameters ($\times$), a prior density fixed to the true value ($\square$), and a fixed-calibration prior scaled by the age of the fossil ($\bigtriangleup$). 
The CP and the precision (the credible-interval width) are shown, respectively, for data generated under different values of $r=\mu/\lambda$: (A \& B) $r=0.1$, (C \& D) $r=0.5$, and (E \& F) $r=0.9$.}
\label{biasedResultsFig1}
\end{figure}

\begin{figure}[tbh!]
\hfil\includegraphics[scale=0.6,angle=0]{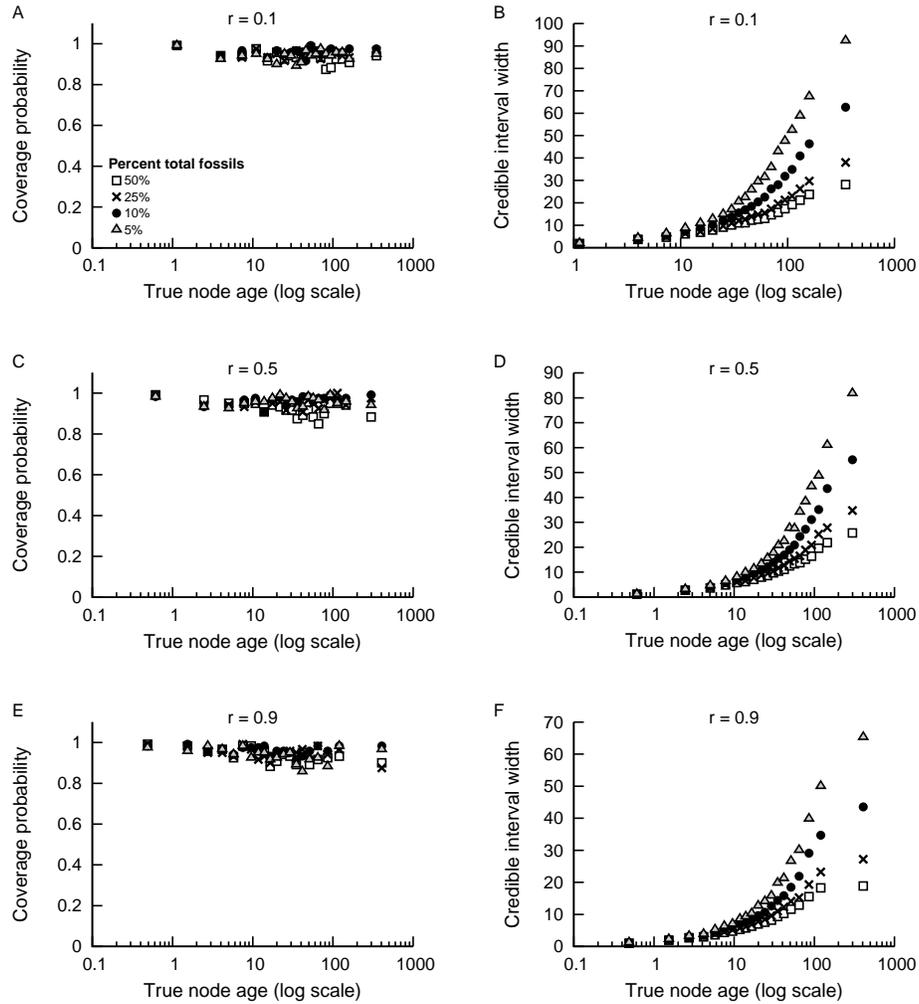}\hfil
\caption{The impact of preservation-biased fossil sampling on the coverage probability and precision (95\% CI width) of node-age estimation under the FBD model.
The results are shown for different percentages of sampled fossils: 5\% (\gtri), 10\% ($\newmoon$), 25\% ($\times$), and 50\% ($\square$). 
Trait-biased fossil sampling was applied to data simulated under different values of $r=\mu/\lambda$: (A \& B) $r=0.1$, (C \& D) $r=0.5$, and (E \& F) $r=0.9$.}
\label{biasedResultsFig2}
\end{figure}

\bigskip
\noindent
{\bf Biased fossil sampling. }
Taphonomic biases -- variation in rates of fossilization, preservation, and recovery -- are prevalent in the fossil record, and these patterns are likely to impact node-age estimates under the FBD model. 
In particular, phylogenetic variation in rates and abilities of preservation is a well known pattern in the tree of life \cite{marshall90,alroy1999,foote1999,wagner2000,foote2001,smith2001,valentine2006}, 
with rich fossil records for some taxonomic groups like Foraminifera \cite{aze2011} or bivalves \cite{valentine2006} and relatively depauperate fossil representation of angiosperms \cite{sanderson2004} and other predominantly soft-tissued organisms \cite{sansom2013}.
To address this factor, we generated sets of non-randomly sampled fossils for all of our simulation replicates (for all values of $r$: 0.1, 0.5, 0.9) by simulating a continuous character for each fossil under a geometric Brownian-motion model.
This model is identical to the autocorrelated log-normal model for relaxing the molecular clock \cite{thorne98,kishino01,thorne02,thorne05} and produces an autocorrelated sampling/preservation rate over the tree with closely related fossils assigned similar rates (see Methods \ref{Pres-Foss-Samp-Meth}). 
We sampled fossils under different values of $\omega$ (5\%, 10\%, 25\%, 50\%) in proportion to their preservation rates, where fossils with high values have a higher probability of being sampled, thus mimicking some degree of trait-based fossil recovery. 
When divergence times were estimated under the FBD model using non-random samples of fossils, the CP and precision were similar to random fossil sampling (Fig.~\ref{biasedResultsFig1} and \ref{biasedResultsFig2}), resulting in patterns much like those presented for randomly sampled fossils in Figures \ref{simFig1} and \ref{simFig2} (and Figs.~\ref{rndTurnoverCompareFig} and \ref{rndTurnoverFossSampFig}). 
Thus, FBD node-age estimates are robust to this type of character-driven fossil sampling. 

\begin{figure}[tbh!]
\hfil\includegraphics[scale=0.5]{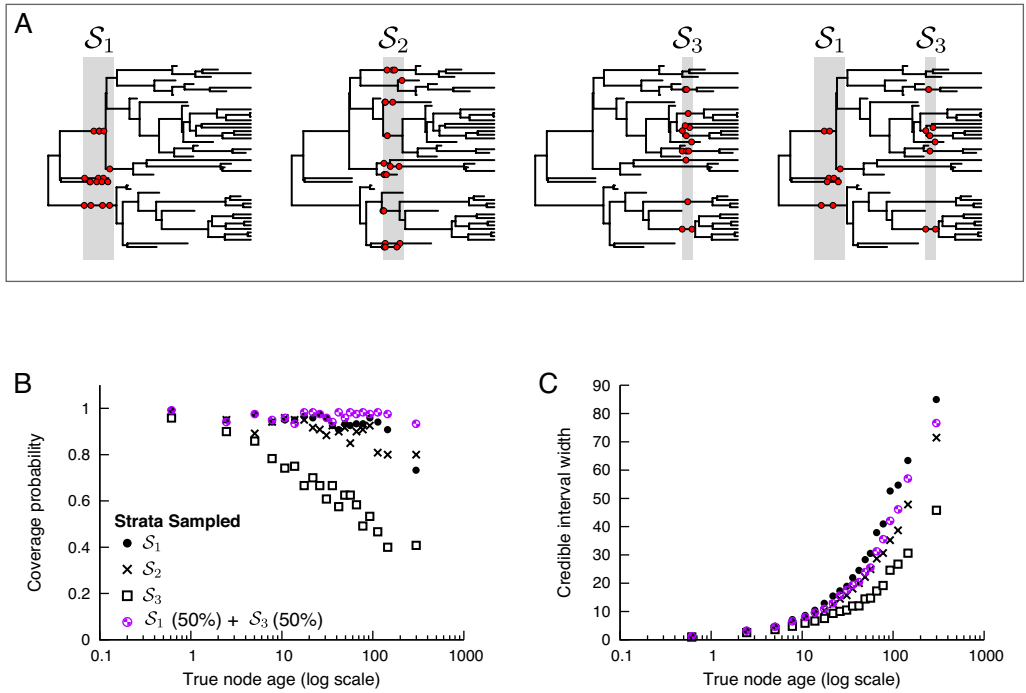}\hfil
\caption{Stratigraphic fossil sampling across three different fossiliferous horizons. (A) An example of the four different sets of fossil samples. Three sets of fossils were assembled by selecting all of the available fossilization events contained within each of the three intervals: $\mathcal{S}_1$, $\mathcal{S}_2$, and $\mathcal{S}_3$. The fourth set of fossils was constructed by sampling 50\% of the fossils in horizon $\mathcal{S}_1$ and half of the fossils in $\mathcal{S}_3$. 
(B) The coverage probability as a function of the true age for estimates using each of the four sets of fossils: $\mathcal{S}_1$ ($\newmoon$), $\mathcal{S}_2$ ($\times$), $\mathcal{S}_3$ ($\square$), and $\mathcal{S}_1+\mathcal{S}_3$ ({\textcolor{purpcol}{$\boldsymbol{\bigotimes}$}}).
(C) The credible interval width for increasing node ages.}
\label{simStratigResultsFig}
\end{figure}

Stratigraphy and gaps in the rock-record \cite{marshall90,marshall1994,marshall1997,weiss1999} are additional factors likely to affect models assuming random and continuous fossil sampling. 
We created three different ``fossiliferous horizons'' for each simulated tree ($r=0.5$ simulations; Methods \ref{Strat-Foss-Samp-Meth}). Each stratum was placed relative to the root so that the midpoints were located at $\mathcal{S}_1 = \frac{3}{4}x_1$, $\mathcal{S}_2 = \frac{1}{2}x_1$, $\mathcal{S}_3 = \frac{1}{4}x_1$. 
The range for each sampling horizon $i$ was $\mathcal{S}_i\pm \frac{1}{10}\mathcal{S}_i$. 
We estimated divergence times by sampling all of the fossils for a single interval at a time, and by assembling a two-strata set of calibration fossils by randomly sampling 50\% of the fossils from $\mathcal{S}_1$ and 50\% from $\mathcal{S}_3$ (Fig.~\ref{simStratigResultsFig}A). 
When fossils are sampled from a single stratum that is relatively close to the present, $\mathcal{S}_3$, the node age estimates have reduced coverage (Fig~\ref{simStratigResultsFig}B) and narrower credible intervals (Fig~\ref{simStratigResultsFig}C), particularly at older nodes, compared with node-age CPs using randomly sampled fossils (as shown in Fig.~\ref{simFig2}). 
This effect is less pronounced when fossils are sampled from intervals closer to the root of the tree, $\mathcal{S}_1$ or $\mathcal{S}_2$, but we still see a marked decrease in CPs when the set of fossils represents a narrow window in time. 
However, when fossils are sampled from two intervals (with half of the fossils sampled from each $\mathcal{S}_1$ and $\mathcal{S}_3$), we find that node age estimates are generally robust to this sampling pattern with high coverage probabilities across the range of node ages (Fig.~\ref{simStratigResultsFig}). 
These results highlight the importance of including fossils spanning the range of the evolutionary history for a given group, yet it is less important to sample continuously throughout that timeframe. 

\begin{figure}[tbh!]
\hfil\includegraphics[scale=0.5,angle=-90]{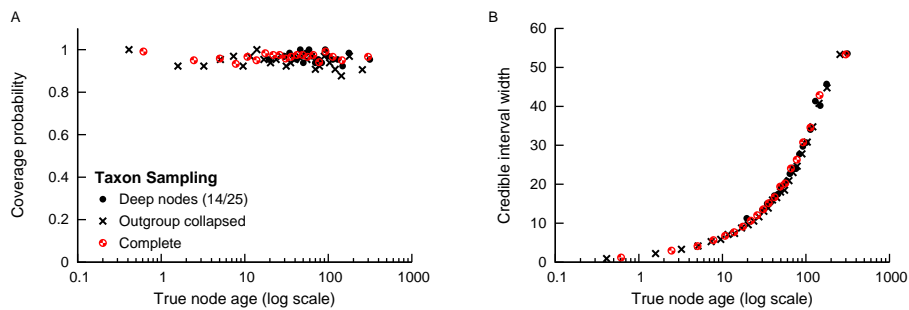}\hfil
\caption{The effect of non-random extant species sampling on the coverage probability and precision of node age estimates under the FBD model with 10\% of the fossils sampled at random.
(A) The coverage probabilities and (B) credible interval widths compared to the true node age when extant species are sampled to maximize the diversity (deep nodes; $\newmoon$), for samples emulating outgroup sampling with significantly imbalanced root nodes ($\times$), and when all extant species are sampled ({\textcolor{red}{$\boldsymbol{\bigotimes}$}}).}
\label{simTaxonSamp}
\end{figure}

\bigskip
\noindent
{\bf Biased extant species sampling. }
In our current implementation of the FBD model, we assume that the probability of sampling extant tips is $\rho = 1$, an unrealistic assumption that is undoubtably violated by most biological datasets. 
We applied two different strategies of non-random taxon sampling (using the $r=0.5$ simulations) to evaluate the robustness of node-age estimates when the $\rho=1$ assumption is explicitly violated.
First, we sub-sampled the extant trees and sequence datasets to mimic taxonomy-driven species sampling, where tips are sampled to maximize the diversity in the tree.
We sampled species to represent the 13 deepest nodes, resulting in trees of 14 extant tips. 
This `diversified' sampling induces trees with long terminal branches \cite{hohna11}.
The second taxon-sampling strategy we evaluated was developed to emulate concentrated ingroup sampling with only 1 or 2 lineages representing outgroups. 
This subsampling resulted in a set of trees with greater imbalance when measured with the weighted-mean imbalance tree-shape statistic \cite{fusco1995,purvis2002,heath08} (Methods \ref{Ext-Spp-Samp-Meth}).
For each simulation replicate and each sampling strategy, we retained all of the fossils in the $\omega_{10}$ set. 
Since some calibration nodes were eliminated by taxon sub-sampling, the fossils calibrating those nodes were assigned to the next ancestral node represented in the tree.
We estimated lineage divergence times under the FBD model 
and found that non-random taxon sampling had no effect on the CPs or 95\% CI widths when compared to the fully sampled trees (Fig.~\ref{simTaxonSamp}). 
These results show estimates of absolute node ages are robust to unequivocal violation of the $\rho=1$ assumption. 

Much like conventional calibration approaches, estimates of absolute node ages under the FBD model are informed by the distribution and sampling of fossils.
However, unlike calibration-density methods, inference is improved as additional fossils are applied to already-calibrated nodes. 
The FBD model is also robust to explicit violation of model assumptions, provided that the fossils are correctly placed and cover a wide range of node ages.
Fundamentally, our results highlight the importance of thorough fossil sampling for robust and accurate node-age estimation.

\subsection{Analysis of Biological Data}

We assembled a phylogenetic dataset of all living bears (Ursidae) plus two outgroup species (\textit{Canis lupus} and \textit{Phoca largha}). 
The sequence data for the complete mitochondrial genomes (mtDNA) and the nuclear interphotoreceptor retinoid-binding protein gene (irbp) were downloaded from GenBank (Table \ref{bearsDNAGBTable}) and aligned using MAFFT \cite{katoh2013}. 
Preliminary analysis of these data using MrBayes v3.2 \cite{Ronquist2012} yielded the same topology (for the overlapping set of taxa) reported in two recent studies \cite{krause2008,dosReis2012}; we then conditioned our divergence time analyses on this topology.

For estimation of absolute speciation times, we compiled a set of fossil ages from the literature (Table \ref{bearFossilTable}). 
This set of fossils included five fossils belonging to the family Canidae (assigned to calibrate the root), five fossils classified as Pinnipedimorpha (assigned to date the MRCA of \textit{P.~largha} and Ursidae), and 14 fossils in the family Ursidae.
Information regarding the phylogenetic placement of the ursid fossils was based on phylogenetic analyses of morphological data \cite{abella12} as well as analyses of mtDNA for extinct Pleistocene subfossils \cite{krause2008}. 
With these data, we estimated divergence times under the FBD model using five stem-fossil ursids, six fossils in the subfamily Ailuopodinae (pandas and relatives), the giant short-faced bear (\textit{Arctodus}; Pleistocene), and two fossil representatives of the genus \textit{Ursus}, including the Pleiocene \textit{U.~abstrusus} fossil, and the Pleistocene cave bear subfossil, \textit{U.~spelaeus} (Table \ref{bearFossilTable}). 
The ages for most fossils are imprecise and therefore represented in the literature as age ranges. 
Because the FBD model assumes that the fossil is associated with a single point in time, we then sampled the age for each fossil from a uniform distribution on its given range (Table \ref{bearFossilTable}). 
Given sufficient fossil sampling, this approach is intended to approximate random recovery.
It would be preferable, however, to treat the ages of fossils as random variables by placing prior densities on fossil occurrence times conditional on their estimated age ranges---this feature is planned for future implementations of the FBD model.  

Using DPPDiv, we estimated absolute species times for all living bears, as well as deep nodes in the caniform tree. 
We used a GTR+$\Gamma$ model of sequence evolution and applied a relaxed-clock model to allow substitution rates to vary across the tree.
The relaxed-clock model we employed assumes that lineage-specific substitution rates are distributed according to a Dirichlet process prior (DPP), such that branches fall into distinct rate categories, yet the number of categories and the assignment of branches to those categories are random variables \cite{heath11}. 
The DPP relaxed-clock model places non-zero prior probability on all 115,975 branch-to-rate configurations including the strict clock (one category), all local molecular clocks, and a model where the rates associated with each branch are independently drawn from a single, underlying distribution (similar to the commonly used, uncorrelated models in the program BEAST \cite{drummond06,Drummond2012}). 
For the FBD model, we assumed putatively noninformative, uniform priors on all rate parameters. 
Additionally, since our simulations demonstrated that node age estimation is robust to extant species sampling, we specified $\rho=1$.
We ran three independent chains, each for 20 million iterations. 
We verified that the three independent runs converged on the same stationary distribution for all parameters using the program Tracer \cite{rambaut09}.
Furthermore, we confirmed that the sequence data were informative on branch rates and node times by comparing the MCMC samples from the three different runs to samples under the prior (\textit{i.e.}, without data).

\begin{figure}[tbh!]
\hfil\includegraphics[scale=0.5]{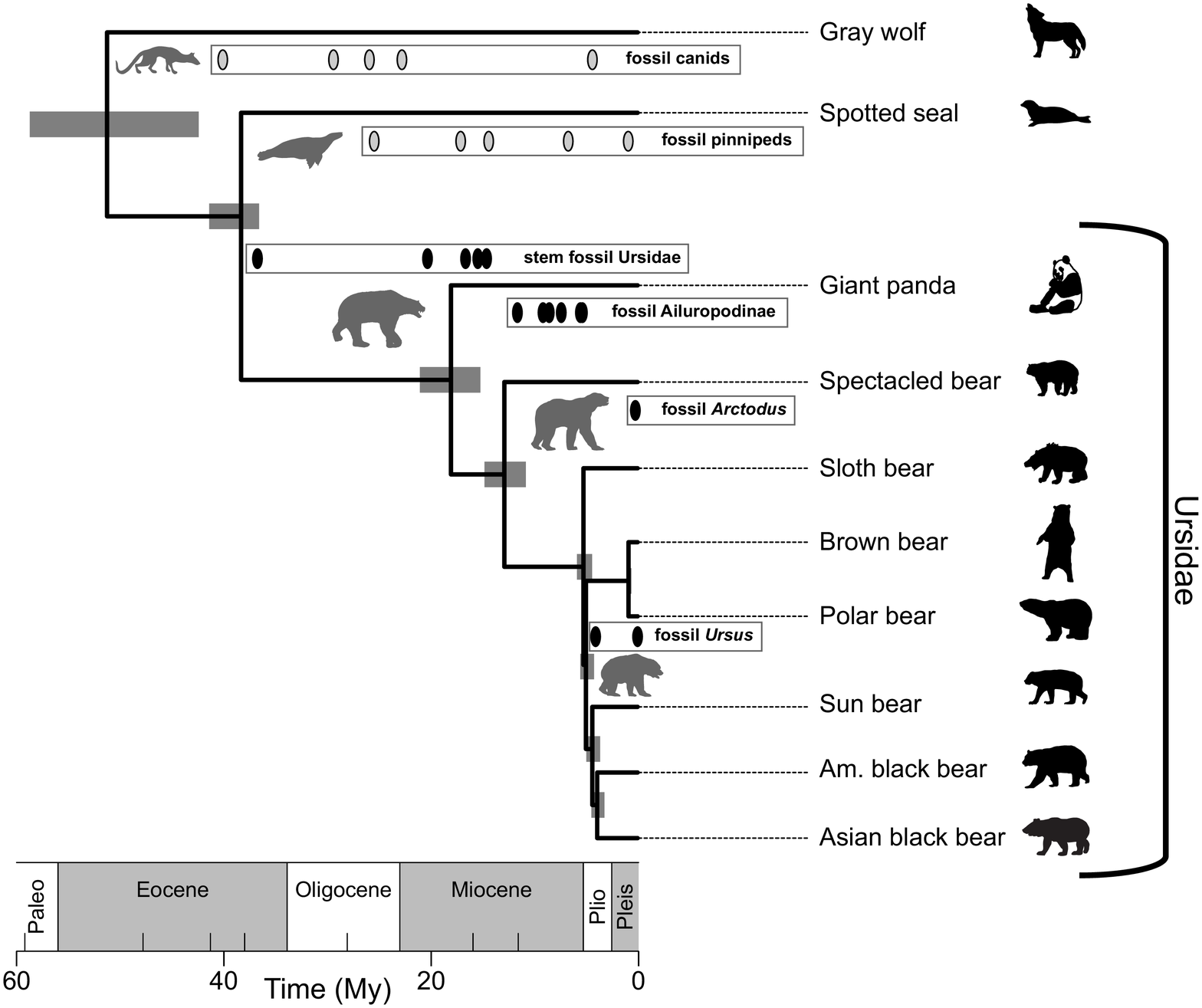}\hfil
\caption{The divergence times of extant bears and two caniform outgroups estimated under the FBD model. 
The branch lengths are in proportion to the mean branch time in millions of years. 
Horizontal node bars represent the 95\% credible interval for node ages. 
The occurrence times for the calibrating fossils are represented within the labeled boxes. 
In each labeled box, the ovals indicate the fossil occurance times. 
The fossils in the family Ursidae are all indicated with black ovals, while the outgroup fossils are shaded light gray.
\textit{Ursus} (including \textit{Melursus} and \textit{Helarctos}) species include the sloth bear, brown bear, polar bear, sun bear, American black bear, and Asian black bear.
(Taxon images available at http://phylopic.org/.)}
\label{bearSummaryFig}
\end{figure}

\begin{figure}[tbh!]
\hfil\includegraphics[scale=0.5]{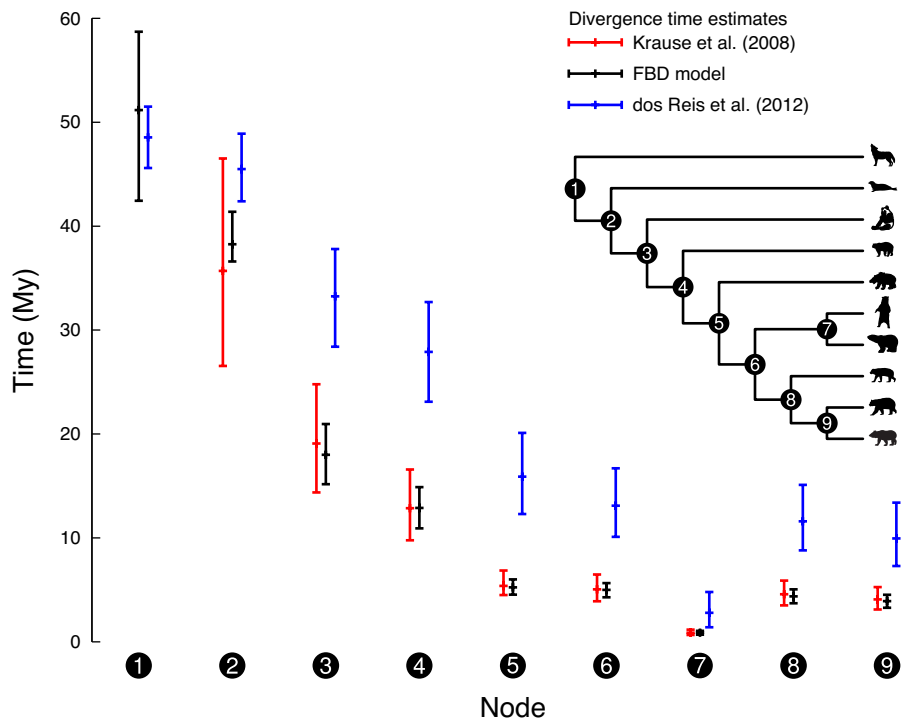}\hfil
\caption{Estimates of bear divergence times under the FBD model compared with time-estimates from previously published studies \cite{krause2008,dosReis2012}. The lines define the lower bound of the 95\% credible set, the mean, and upper bound of the 95\% credible set for each node in the phylogeny labeled $1,\ldots,9$, starting with the root. 
Estimates under the FBD model are shown in black, node ages from Krause et al. \cite{krause2008} are shown in red, and ages estimated by dos Reis et al. \cite{dosReis2012} are shown in blue. }
\label{bearCompareDivtimeFig}
\end{figure}

We summarized the divergence times sampled by MCMC from the three independent runs using tools in the Python library DendroPy \cite{sukumaran10}. 
The divergence times of all living bears and occurrence times of calibrating fossils are summarized in Figure \ref{bearSummaryFig}. 
On average, the node ages estimated under the FBD model are consistent with the ages estimated by Krause et al.~\cite{krause2008}, with overlapping 95\% CIs for all common nodes (Fig.~\ref{bearCompareDivtimeFig}), suggesting the radiation of the genus \textit{Ursus} occurred in the late Miocene to mid Pliocene (Fig.~\ref{bearSummaryFig}).
In contrast, dos Reis et al.~\cite{dosReis2012} uncovered much older ages for all nodes except for the node representing the MRCA of Canidae and Ursidae (the root in Fig.~\ref{bearSummaryFig}). 
Their results indicate a much earlier origin of crown ursids, with much of the diversification of present-day \textit{Ursus} occurring in the mid-to-late Miocene. 
Our analyses were most similar to those of Krause et al.~\cite{krause2008}; they estimated divergence times using mitochondrial genomes under a global molecular clock.
This gene region dominated our alignment, and analyses under the DPP model indicate low among-lineage substitution-rate variation with a median of two rate categories (95\% credible interval: 1--4). 
Furthermore, although they used calibration densities, the common fossils, \textit{U.~abstrusus} and \textit{Parictis montanus}, between our study and Krause et al.~\cite{krause2008} appear to strongly influence absolute node-ages. 
In contrast, dos Reis et al.~\cite{dosReis2012} estimated divergence times for 274 mammals using a two-step approach by first estimating the node ages for 36 species using nuclear genomes, then using the posterior estimates to inform analyses of 12 mitochondrial protein genes for all 274 species. 
Their analysis of the 36-taxon tree included only two carnivores, \textit{Canis familiaris} and \textit{Felis catus}, and thus the age estimates of the bears in the 274 species tree were uncalibrated, yet informed by the ages of all other nodes in the mammal tree. 

The dissimilar divergence-time estimates resulting from distinctly different Bayesian methods and datasets reveal that the speciation times within the caniforms may still be an open question.
Moreover, elucidating the time-calibrated phylogeny of this group calls for a more comprehensive approach that includes more species and sequence data (as in \cite{dosReis2012}) in conjunction with a mechanistic diversification model that allows for inclusion of the rich fossil history of the group.

\section{Conclusions}
We introduced a new macroevolutionary model that integrates fossil information into the lineage-based diversification model and overcomes many of the limitations of calibration-density methods.
Standard calibration-density approaches using fossils have inherent flaws that can lead to biased estimates of node ages and measures of uncertainty, particularly when applied using fixed parameters \cite{heath12}. 
Fundamentally, the prior densities used to calibrate nodes are not derived from any underlying biological process but are instead intended to characterize the biologist's uncertainty in the age of the calibrated node with respect to its oldest descendant fossil. 
As a result, estimates of absolute speciation times are driven by the choice and parameterization of these calibration densities. 
The fossilized birth-death process considers calibrating fossils and extant species as part of a unified diversification process and provides a more mechanistic model for lineage divergence times.

Our simulations show that estimates under the FBD model have higher coverage when compared with the most robust calibration approaches. 
This result holds even relative to estimates using calibration densities centered on the true value (an ideal, but unrealistic scenario). In particular the FBD model also provides reliable estimates when assumptions of the FBD model are violated in the data, namely, biased sampling of fossils or extant species --- scenarios which are probably very common.
Importantly, since increasing the number of calibrating fossils results in a corresponding increase in the precision of node age estimates, the FBD model is better at capturing our statistical uncertainty in the timing of speciation events. 
By contrast, the precision of node age estimates under calibration-density approaches is entirely controlled by the precision of the prior distributions on calibrated nodes, this is particularly the case for fixed calibration densities.
Thus, the FBD model eliminates one of the greatest challenges imposed on biologists applying Bayesian divergence-time methods: choosing and parameterizing calibration densities for multiple nodes. 

Phylogenetic analysis under a unified model of diversification employs all of the fossils available for a group. 
This feature represents another significant advantage over traditional approaches that condense all of the fossil information associated with a given node to a single minimum age estimate. 
Our results highlight the importance of thorough fossil sampling, therefore every reliably identified and dated fossil species is useful and may improve estimation of both node ages and parameters of the diversification model. 
In fact, inclusion of fossil lineages in macroevolutionary studies leads to more accurate inferences of patterns of speciation and extinction \cite{Quental2010} and rates of phenotypic-trait evolution \cite{slater2012}.
This factor underscores the importance of carefully curated and analyzed paleontological collections, thus motivating collaboration with experts on extinct organisms and critical assessment of fossil specimens \cite{Parham2012}. 
Moreover, comprehensive models like the FBD have the potential to address interesting questions about diversity through time and harness the wealth of information available in online databases like the Paleobiology Database (\href{http://paleodb.org}{http://paleodb.org}). 

Alternative sources of calibrating information are often applied to date species phylogenies, particularly when fossil information is unavailable. 
Therefore, it is important to note that the FBD model is explicitly for \textit{fossil} calibration.
Non-fossil calibration times are typically derived from biogeographical dates or node-age estimates from previous studies (\textit{e.g.}, secondary calibrations). 
Thus, the FBD model is not an appropriate prior for calibration with these data and calibration-density approaches are still necessary.
For these analyses, hierarchical calibration models \cite{heath12} and methods that condition on node calibrations \cite{yang06,heled12} 
are strongly recommended.

The fossilized birth-death model and its parent model, the serially sampled birth-death process, offer a rich basis for the development of complex, biologically informed models of macroevolution that incorporate both modern and fossil species. 
Improving the integration of fossil data with extant-species data in a phylogenetic framework is an important step toward enhancing our understanding of evolution and biodiversity.
As morphological datasets uniting fossil and modern species become available, methods combining the SSBD and FBD models will be necessary to answer macroevolutionary questions in a phylogenetic context. 
New models based on these processes can additionally account for our prior knowledge of taphonomic controls and the geologic record. 
In fact, extensions of the SSBD model that accommodate variation in diversification and sampling parameters are already useful for analysis of infectious disease data \cite{Stadler012,StadlerYang2013,Stadler2013,stadlerbonhoeffer2013}. 
The phylogenetic patterns of rapidly evolving viruses modeled by these processes are analogous to macroevolutionary patterns of lineage diversification. 
Mechanistic diversification models that account for biological factors and properties of the fossil and geologic records can provide a better understanding of lineage diversification across the tree of life.

\FloatBarrier
\section{Methods}
\subsection{The Probability of a FBD tree}\label{FBDprobSec}
Stadler \cite{stadler10} defines the probability of an explicit phylogenetic tree of extant species and fossils -- \textit{i.e.}, a sampled tree -- under the serial-sampled birth-death (SSBD) model (Fig.~\ref{fbdTree}A).
The fossilized birth-death (FBD) model extends from this previous work and we define the probability of a \textit{FBD tree}. 
The FBD tree is a sampled tree where the precise phylogenetic topology of any fossil $f$ attaching to the sampled tree is not specified, only the fossil's ancestral node, its age $y_f$, and the time $z_f$ at which the fossil attaches to the tree are evaluated (Fig.~\ref{fbdTree}B).

For our FBD calibration method, we calculate the probability of a FBD tree: $f[\mathcal{T} \mid \lambda, \mu, \psi, \rho, x_{1} ]$, where the probability of the topology, internal node ages, and fossil attachment times ($\mathcal{T}$) for $n$ extant species and $m$ calibrating fossils is conditional on the hyperparameters of the FBD model ($\lambda$, $\mu$, $\psi$, $\rho$) and the age of the root node ($x_1$). 
To state $f[\mathcal{T} \mid \lambda, \mu, \psi, \rho, x_{1} ]$ we define additional notation (Table \ref{FBDMathNotation}).
Let ${\cal V}$ be the set of internal node indices in the extant species phylogeny, ${\cal V} = (1,\ldots,n-1)$, labeled in preorder sequence such that $1$ is the index representing the root of the tree;
and the age for any internal node $i$ is $x_i$ (for $i \in {\cal V}$). 
The occurrence times and ancestral nodes in the extant tree are provided for $m$ sampled fossil specimens. 
For any given FBD tree, $k$ of $m$ fossils lie directly on branches in the sampled tree and are therefore ancestor fossils.
Accordingly, $m-k$ of $m$ fossils attach to the sampled tree at a speciation event and thus induce unobserved lineages. 
We denote the vector of fossil calibration indices as 
${\cal F} = (1,\ldots,m)$; $y_f$ is the age of fossil $f$ obtained from the fossil record and $z_f$ is the time at which the fossil attaches to the extant tree (for $f \in {\cal F}$). 
Under the FBD model the probability of the sampled tree is invariant to changing the attachment lineage of fossil $f$ at time $z_f$, therefore we multiply the sampled tree probability by the number of possible attachment lineages, $\gamma_f$ (\textit{e.g.}, for fossil 1 in Fig.~\ref{fbdTree}B there are two attachment lineages, $\gamma_1 = 2$). 
Moreover, since the fossil may attach to the ``right'' or ``left'' in the sampled tree (as our tree probability is on so-called `oriented trees' \cite{Ford2009}), we we additionally multiply each $\gamma_f$ by a factor of $2$.
Finally, if the fossil is an ancestral fossil, we do not account for the speciation event on the sampled tree, thus
${\cal I}_f$ is the indicator function for fossil $f$, where ${\cal I}_f = 0$ if the fossil is ancestral, otherwise ${\cal I}_f = 1$.

Given the notation defined above, the probability of a FBD tree is
\begin{eqnarray} \label{EqnTreeLik}
f[\mathcal{T} \mid \lambda, \mu, \psi, \rho, x_{1} ]&=& \frac{1}{(\lambda(1-\hat{p}_0(x_{1})))^2} \frac{4\lambda \rho}{q(x_{1})} \prod_{i \in {\cal V}} \frac{4 \lambda \rho}{q(x_i)} \prod_{f \in {\cal F}} \psi \left(2 \gamma_f \lambda  \frac{p_0(y_f)q(y_f)}{q(z_f)}\right)^{{\cal I}_f},
\end{eqnarray}
with
\begin{eqnarray*}
c_1 &=& |\ \sqrt{(\lambda - \mu - \psi)^2 + 4\lambda\psi}\ |,\\
c_2 &=& - \frac{\lambda - \mu - 2\lambda\rho-\psi}{c_1},\\
q(t) &=& 2(1-c_2^2) + e^{-c_1 t} (1-c_2)^2 + e^{c_1 t}(1 + c_2)^2,\\
p_0(t) &=& 1+ \frac{-(\lambda - \mu - \psi) + c_1 \frac{e^{-c_1 t}(1-c_2)-(1+c_2) }{e^{-c_1 t} (1-c_2)+(1+c_2)}} {2 \lambda},\\
\hat{p}_0(t) &=& 1-\frac{\rho (\lambda-\mu)}{\lambda \rho + (\lambda (1-\rho) - \mu) e^{-(\lambda-\mu)t}}. 
\end{eqnarray*}
We note that $p_0(t)$ is the probability that a lineage at time $t$ in the past has no sampled extant species or sampled fossil descendants,  $\hat{p}_0(t)$ is the probability that a lineage at time $t$ in the past has no sampled extant species descendants and arbitrarily many fossil descendants. 
Further, we define $p_1(t)$ as the probability that an individual alive at time $t$ before the present has precisely one sampled extant descendant and no sampled fossil descendants \cite{stadler10}, and we have $p_1(t)= \frac{4\rho}{q(t)}$. 
Alternatively, we can calculate the probability of a FBD tree conditional on the stem age $x_0$. 
This parameterization is useful when stem fossils are available for branches not included in the extant tree. 
We state this alternative probability in Section \ref{SecStem}.

The four FBD model parameters (\textit{e.g.}, $\lambda,\mu,\psi,\rho$) are required to write down the probability of a FBD tree, while for $\psi=0$ only two parameters (\textit{e.g.}, $\lambda-\mu$ and $\lambda \rho$) are required (Section \ref{SecFBDepi}). 
Therefore, when using fossils, we expect to be able to infer all four parameters for sufficiently large datasets, whereas with extant species data alone, it is not possible to separately estimate $\lambda,\mu,\rho$. 


\subsubsection{Probability of FBD Tree Conditioned on the Stem Age}\label{SecStem}
For some datasets, we may wish to condition on the age of the stem $x_0$ instead of the age of the crown (root) node $x_1$. 
This approach is typically applied to analyses of infectious diseases, in the presence of reliable information about the origin times of an epidemic. 
For species-level data, however, it is difficult to formulate a prior distribution on the age of the stem node without inclusion of an outgroup or fossil information.
When stem fossils are available, they may be applied to the stem lineage without including any outgroup species. 
The probability of a FBD tree with stem age $x_{0}$ is
\begin{eqnarray}
f[\mathcal{T} \mid \lambda, \mu, \psi, \rho, x_{0} ]&=& \frac{1}{\lambda(1-\hat{p}_0(x_{0}))} \frac{4 \lambda \rho}{q(x_{0})} \prod_{i \in {\cal V}} \frac{4 \lambda \rho}{q(x_i)} \prod_{f \in {\cal F}} \psi \left(2 \gamma_f \lambda  \frac{p_0(y_f)q(y_f)}{q(z_f)}\right)^{{\cal I}_f}.
\end{eqnarray}
See Table \ref{FBDMathNotation} for notation definitions.
Under this formulation of the model, one may assign fossils to the stem of the tree.

\subsubsection{Probability of a FBD Tree Without Sampled Ancestors}\label{SecFBDepi}
Consider a modified FBD model where, upon fossil recovery (with rate $\psi$) the lineage dies; we call this model the FBDd model. The probability of a FBDd tree directly follows from Equation (\ref{EqnTreeLik}) together with  \cite{Stadler012},
\begin{eqnarray}
f[\mathcal{T} \mid \lambda, \mu, \psi, \rho, x_{1},k=0]&=& \frac{1}{(\lambda(1-\hat{p}_0(x_{1})))^2} \frac{4\lambda \rho}{q(x_{1})} \prod_{i \in {\cal V}} \frac{4 \lambda \rho}{q(x_i)} \prod_{f \in {\cal F}}\psi \left(2 \gamma_f \lambda \frac{q(y_f)}{q(z_f)}\right). \label{EqnFBDepi}
\end{eqnarray}
This model has been used when applying phylogenetic methods to pathogen sequence data \cite{Stadler012,Stadler2013}.
It was shown in previous papers \cite{stadler09,Stadler2013} that for either $\rho=0$ or $\psi=0$ and using $p_0(t)$ instead of $\hat{p}_0(t)$, the tree probability only depends on two out of the three parameters $\lambda$, $\mu$ and either $\rho$ or $\psi$. More generally, we can directly observe from Equation \ref{EqnFBDepi} that, when using $p_0(t)$ instead of $\hat{p}_0(t)$, the probability does not require four  parameters $\lambda,\mu,\psi,\rho$, but instead can be written as a function of three parameters  $\lambda-\mu-\psi$, $\lambda \rho$, $\lambda \psi$.

\subsubsection{Parameter Correlations}\label{SecFBDParCorr}
The parameter correlations discussed for the FBDd model (Section \ref{SecFBDepi}) disappear in the FBD model. Equation (\ref{EqnTreeLik}) involves the four parameters ($\lambda,\mu,\psi,\rho$) and these parameters each contain information:
 $c_1$ depends on two parameters $\lambda - \mu - \psi$ and $\lambda \psi$. 
Additionally, $c_2$ depends on  $\lambda \rho$. We can write all terms of the likelihood using the previous three parameters, except for 
 $\psi^k  \prod_{f \in {\cal F}} p_0(z_f)^{{\cal I}_f}$ which depends on $\lambda$ and $\psi$ individually.  
 Setting $\rho=1$, as we do in all of our analyses, still leads to  three degrees of freedom (i.e. three  parameters are required to write down the tree probability).

\subsection{MCMC Approximation under the FBD model}\label{MCMCMethodSec}
We implemented the FBD model in a Bayesian framework for estimating species divergence times on a fixed tree topology by building the model into an existing program, DPPDiv (version 1.1; available at \href{https://github.com/trayc7/FDPPDIV}{https://github.com/trayc7/FDPPDIV}) \cite{heath11,heath12,Darriba2013}. 
The input data ($\mathcal{D}$) are:
\begin{center}
\begin{tabular}{rcl}
$\tau$ & \hspace{6mm} & Extant species tree topology\\
$X$ & & DNA sequences for $n$ extant species\\
$\boldsymbol{y} = (y_1,\ldots,y_m)$ & & Vector of occurrence times for $m$ fossils\\
$\mathcal{C}$ & & Known ancestral nodes in $\tau$ for all fossils.\\
\end{tabular} 
\end{center}
Note that these are the same input data required for divergence-time estimation using calibration density approaches \cite{yang06,ho09,heath12}. 
We use MCMC to approximate the joint posterior distribution of internal node ages and fossil attachment times ($\mathcal{T}$) together with all other model parameters. 

The FBD model acts as a prior on speciation times, thus explicitly assuming that this model generated $\mathcal{T}$. 
Additionally, we assume prior distributions on parameters of the model of sequence evolution and on the substitution rates associated with each branch in the tree.
These models correspond to the exchangeability rates and nucleotide frequencies of the GTR model, the shape parameter of the gamma distribution on site rates, and the parameters of the relaxed clock model describing lineage-specific rate variation across the tree. 
The implementation of these models are described in detail in Heath et al. \cite{heath11} and we use the notation $\theta$ to represent these parameters. 
Using this framework, we sample from the posterior distribution of FBD trees:
$$f[\mathcal{T},\lambda,\mu,\psi,\rho,\theta, x_1|  \mathcal{D}=(\tau, X, \mathcal{C}, \boldsymbol{y})] = \frac{f[X|\mathcal{T},\theta] f[\mathcal{T}|\lambda,\mu,\psi,\rho,x_{1}] f[\lambda,\mu,\psi,\rho,x_{1},\theta]} 
{f[\mathcal{D}]}.$$
We note that 
$f[X|\mathcal{T},\theta]$ is the likelihood of the tree and sequence model parameters, which we calculate with Felsenstein's pruning algorithm \cite{felsenstein73,felsenstein81},
$f[\mathcal{T}|\lambda,\mu,\psi,\rho,x_{1}]$ is the prior probability of the FBD tree given in Equation \ref{EqnTreeLik}, and $f[\lambda,\mu,\psi,\rho,x_{1},\theta]$ is the prior distribution on the model parameters and hyperparameters (specified by the user). 
Because we use the numerical method MCMC, we avoid computation of the normalizing constant $f[\mathcal{D}]$.

Our implementation of the FBD model assumes that $\mu\leq\lambda$, otherwise the process will go extinct with probability $1$. 
Furthermore, instead of the parameters $\lambda$, $\mu$, and $\psi$, we use the following parameterization:
\begin{center}
\begin{tabular}{rcl}
$d=\lambda-\mu$ & \hspace{6mm} & Net diversification rate\\
$r=\mu / \lambda$ & & Turnover\\
$s=\psi / (\mu+\psi)$ & & Probability of fossil recovery prior to species extinction.\\
\end{tabular} 
\end{center}
Importantly, we can recover $\lambda$, $\mu$, and $\psi$ via: 
$$\lambda=\frac{d}{1-r}, \quad \mu=\frac{rd}{1-r}, \quad \psi = \frac{s}{1-s} \frac{rd}{1-r}.$$
The $d,r,s$ parameterization has the advantage that $r,s,\rho \in [0,1]$  and only $d$ 
is on the interval $(0,\infty)$, whereas the parameterization using $\lambda,\mu,\psi$ requires a prior on an unbounded interval $(0,\infty)$ for each of the three parameters.

We employed standard MCMC proposals for operating on the parameters and hyperparameters of the FBD model ($d,r,s,\rho$, and $x_{1}$) and sequence substitution model ($\theta$), changing the ages ($x$) of internal nodes, and updating the attachment ages of the fossils ($z$).
These proposal mechanisms are described in greater detail in previous implementations of Bayesian inference software \cite{thorne98,huelsenbeck00a,huelsenbeck01c,thorne02,drummond06,yang06,huelsenbeck07b,heath11,Ronquist2012}, and were all previously available in DPPDiv.

We formulated new reversible-jump Markov chain Monte Carlo (rjMCMC) proposals to sample fossil-attachment configurations and determine if fossils lie directly on lineages (ancestral fossils) or represent `tip fossils' by attaching to the sampled tree via speciation (Fig.~\ref{fbdTree}B). 
Moves that cause a fossil to become ancestral to an extant species, as well as reciprocal moves changing an ancestral fossil to one that forms an extinct lineage, result in a dimensionality change to the FBD tree by altering the number of speciation events, 
therefore, rjMCMC \cite{green95} proposals are needed to sample from the posterior distribution.  
Below (Section \ref{SI-Sec-rjMCMCmoves}), we outline our rjMCMC proposals on the fossil-attachment configurations.
These proposals, which are similar to the polytomy proposals of Lewis et al.~\cite{Lewis2005}, result in a probability mass for each fossil $f$ on the state where $y_f = z_f$. 
Ultimately, our MCMC framework allows us to sample ages of nodes in the extant tree while marginalizing over the fossil attachment times and FBD model hyperparameters. 

\subsubsection{New rjMCMC Proposals for Sampling FBD trees}\label{SI-Sec-rjMCMCmoves}
Using MCMC to sample FBD tree configurations requires necessary dimensional changes to move between a state where a given fossil lies directly on a branch in the tree (ancestral fossil) to one where that fossil forms its own, unobserved lineage (Fig.~\ref{fbdTree}B). 
Green \cite{green95} described an extension of MCMC -- reversible-jump Markov chain Monte Carlo (rjMCMC) -- that allows sampling over parameter spaces when the dimensionality or number of parameters is unknown. 
When a calibrating fossil represents an ancestor of a lineage in the FBD tree, the attachment age is not considered in the calculation of the probability (Eq.~\ref{EqnTreeLik}) because the fossil does not originate from a new speciation event. 
If we wish to propose a new state where the `ancestral fossil' becomes a `tip fossil' and arises via speciation creating a new node at its attachment time, then we must consider the dimensionally change as this fossil will now be considered in the FBD tree likelihood. 

Our rjMCMC implementation involves two moves: (1) a move that adds a branch to the tree---\textit{add-branch}---by changing an ancestral fossil to a tip fossil and (2) a move that proposes the deletion of a branch in the FBD tree---\textit{delete-branch}---by converting a tip fossil to one that is ancestral. 
These moves are analogous to the reversible-jump implementation of polytomy proposals described in Lewis et al.~\citet{Lewis2005}. 
For this rjMCMC proposal, we chose add-branch with probability $\alpha$ and delete-branch with probability $1-\alpha$, where $\alpha=0.5$. 
Any rjMCMC move requires that the acceptance probability includes a Jacobian term that corrects for the change in the number of parameters, along with the likelihood ratio, prior ratio, and Hastings ratio \cite{green95}.
Note that these moves rely on the values $m$ and $k$, where $m$ is the total number of fossils and $k$ is the number of $m$ fossils that are ancestral.

\paragraph{Add-Branch Move: Operations, Hastings Ratio, and Jacobian ---}\label{SI-Sec-AddBranch}
If the add-branch move is selected for fossil $f$, then this proposes a state with an additional parameter, $z_f$. 
We must account for the reciprocal move---the delete-branch move that will reverse the proposed add-branch move on fossil $f$---to calculate the Hastings ratio.
Following the description in Lewis et al.~\cite{Lewis2005}, we describe the steps necessary to (A) propose the add-branch move and (B) propose the corresponding delete-branch move and calculate the Hastings ratio and Jacobian term.
The operations all rely on the probability $\alpha$ of choosing the add-branch move, which in our implementation is $\alpha=0.5$.

\bigskip
\noindent\textbf{\textit{(A) The operations necessary for proposing the add-branch move}}
\begin{enumerate}
\item Choose add-branch rather than delete-branch move (probability $g$), where 
\[
    g= 
\begin{cases}
    \alpha,& \text{if } 0<k<m\\
    1.0,& \text{if } k=m\\
    0,              & \text{otherwise } (k=0). 
\end{cases}
\]
\item Choose an ancestral fossil $f$, with age $y_f$ and calibrates node $i$ (probability $1/k$).
\item Choose branch length $\nu^*$ from a uniform distribution on $(0,x_i-y_f)$ where $x_i$ is the age of node $i$, which is the ancestor of fossil $f$. Drawing from a uniform distribution gives $\nu^* = (x_i-y_f)u$, where $u$ is a uniform random variable on the interval (0,1). (The uniform distribution has probability density 1.)
\end{enumerate}
After these three steps are performed, the proposed number of ancestral fossils is updated to $k^*=k-1$.

\bigskip
\noindent\textbf{\textit{(B) The operations necessary for proposing the reciprocal delete-branch move}}
\begin{enumerate}
\item Choose delete-branch rather than add-branch move (probability $h$), where
\[
    h= 
\begin{cases}
    \alpha,& \text{if } 0<k^*<m\\
    1.0,& \text{if } k^*=0.\\
\end{cases}
\]
\item Choose a tip fossil $f$ (probability $1/(m-k+1)$).
\end{enumerate}

\paragraph*{\textit{Hastings ratio ---}}
The Hastings ratio for the add-branch move is the probability of the reverse move (B) divided by the probability of the add-branch move (A). 
This simplifies to: 
$$\phi_a \frac{k}{m-k+1},$$
where $\phi_a={1/\alpha}$ if proposed tree has no ancestral fossils, $\phi_a=\alpha/1$ if current tree has no fossil tips, and $\phi_a=1$ otherwise.
More explicitly, if $\alpha=0.5$, then $\phi_a$ is equal to
\[
    \phi_a = \frac{h}{g}= 
\begin{cases}
    0.5,& \text{if } k=m \text{ and } k^*>0\\
    2.0,& \text{if } k^*=0\\
    1,              & \text{otherwise. } 
\end{cases}
\]

\paragraph*{\textit{Jacobian ---}}
Since this move adds a parameter to the FBD tree, the Jacobian term corrects for that change in dimensionally and is, simply:
$$ \left|\frac{\partial \nu^*}{\partial u} \right|= x_i-y_f.$$
Note that $\nu^*$ is the length of the new branch, $u$ is the uniform-random deviate, $x_i$ is the age of the node calibrated by fossil $f$, and $y_f$ is the age of $f$. 

\paragraph{Delete-Branch Move: Operations, Hastings Ratio, and Jacobian ---}\label{SI-Sec-DelBranch}
If fossil $f$ is currently a tip fossil, arising from a speciation event and attaching to the FBD tree along an extinct branch, we propose the delete-branch move to change to a state where $f$ is an ancestral fossil, thus removing the induced speciation event at time $z_f$.
Proposing a delete-branch move requires (A) calculating the probability of proposing to delete a branch and (B) the probability of proposing the reciprocal add-branch move. 
Additionally, like the add-branch described above (Section \ref{SI-Sec-AddBranch}), the delete-branch move relies on the probability $\alpha$ of choosing the add-branch move. 

\bigskip
\noindent\textbf{\textit{(A) The operations necessary for proposing the delete-branch move}}
\begin{enumerate}
\item Choose delete-branch rather than add-branch move (probability $g$), where
\[
    g= 
\begin{cases}
    \alpha,& \text{if } 0<k<m\\
    1.0,& \text{if } k=0\\
    0,              & \text{otherwise } (k=m). 
\end{cases}
\]
\item Choose a tip fossil $f$ (probability $1/(m-k)$).
\end{enumerate}
These steps update the number of ancestral fossils $k$ to $k^*=k+1$.

\bigskip
\noindent\textbf{\textit{(B) The operations necessary for proposing the reciprocal add-branch move}}
\begin{enumerate}
\item Choose add-branch rather than delete-branch move (probability $h$), where
\[
    h= 
\begin{cases}
    \alpha,& \text{if } 0<k^*<m\\
    1.0,& \text{if } k^*=m.\\
\end{cases}
\]
\item Choose an ancestral fossil $f$, with age $y_f$ and that calibrates node $i$ (probability $1/(k+1)$).
\item Choose branch length $\nu^*$ from a uniform distribution on $(0,x_i-y_f)$ where $x_i$ is the age of node $i$, which is the ancestor of fossil $f$. Drawing from a uniform distribution gives $\nu^* = (x_i-y_f)u$, where $u$ is a uniform random variable on the interval (0,1). (The uniform distribution has probability density 1.)
\end{enumerate}

\paragraph*{\textit{Hastings ratio ---}}
The Hastings ratio for the delete-branch move is the probability of the reverse move (B) divided by the probability of the delete-branch move (A), which gives: 
$$\phi_d \frac{m-k}{k+1}$$
where $\phi_d=\alpha/1$ if current tree has no ancestral fossils, $\phi_d=1/\alpha$ if proposed tree has no fossil tips, and $\phi_d=1$ otherwise.
Specifically, if $\alpha=0.5$, then $\phi_d$ is equal to
\[
    \phi_d = \frac{h}{g}= 
\begin{cases}
    0.5,& \text{if } k=0 \text{ and } m>1\\
    2.0,& \text{if } k^*=m\\
    1,              & \text{otherwise. } 
\end{cases}
\]

\paragraph*{\textit{Jacobian ---}}
The dimension change of the delete-branch move involves reducing the number of speciation events by transforming a tip fossil to an ancestral fossil. 
Thus, the Jacobian term for this move is:
$$ \left|\frac{\partial u}{\partial \nu^*} \right| = \frac{1}{x_i-y_f}. $$

\bigskip
\subsection{Simulation Study: Data Generation}\label{MethSimDetails}

\subsubsection{Trees, Fossils, and Sequences}
We evaluated the performance -- accuracy, precision, and robustness -- of absolute node-age estimation under the FBD model using simulated data.
Complete tree topologies and branch times were generated under a constant-rate birth-death process conditional on $n=25$ extant species using the generalized sampling approach \citep{hartmann10,stadler11} (simulation source code available at \href{https://github.com/trayc7/FossilGen}{https://github.com/trayc7/FossilGen}). 
Three separate sets of simulated trees were generated, each with 100 replicates, such that the turnover rate ($r = \mu/\lambda$) varied between the sets: (A) $r = 0.1$, (B) $r = 0.5$, and (C) $r = 0.9$. 
The rate of diversification ($d = \lambda - \mu$) was adjusted so that the expected root age ($x_1$) was approximately equal to 200: (A) $d = 0.0134$, (B) $d = 0.0106$, and (C) $d =0.0041$.

\begin{figure}[tbh!]
\hfil\includegraphics[scale=0.5,angle=0]{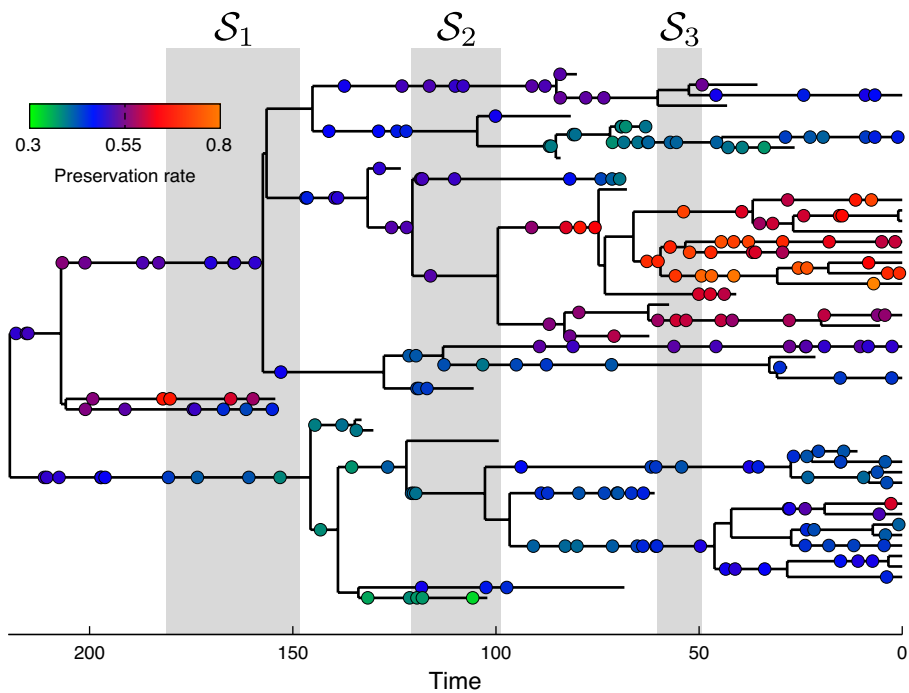}\hfil
\caption{Example of a simulated birth-death tree with fossilization events (SSBD tree). 
This tree summarizes each of the different sampling conditions applied in this study.
The colors indicate the preservation rate used for trait-biased fossil sampling, ranging from green (low rate) to orange (high rate). The sampling rate parameter was generated under a geometric Brownian motion model, such that closely related fossils have similar rate values. Three stratigraphic intervals were also imposed on the tree, each placed relative to the root. The various fossil sampling strategies employed in this study were based on this absolute fossil history.}
\label{simulatorExample}
\end{figure}

An absolute fossil history was generated on each complete phylogeny according to a Poisson process with rate $\psi = 0.1$.
At this step in our forward-time simulation model, the Poisson rate, $\psi$, represents the rate of fossilization opportunity over the tree, thus this set of fossils is the complete fossil record without accounting for preservation and recovery. 
The complete tree with absolute fossil history corresponds to a simulation under the serially sampled birth-death (SSBD) process \cite{stadler10} (Fig.~\ref{simulatorExample}). 
Trees generated under this model with $\psi = 0.1$ produced dense fossil records for each of our simulations (Table \ref{simFossNumsTable}). 
We chose to vary the turnover parameter, $r$, because this value controls the density of fossils. 
Under high values of $r$, there are more lineages on which fossilization events can occur, resulting in more sampled fossils on these trees.
Conversely, trees with low turnover will produce fewer fossils, with a greater proportion of fossils lying directly on branches of the extant tree.

DNA sequence data were generated for every extant taxon across all simulation replicates.
First, we sampled a relative clock rate for each extant tree from a gamma distribution with a shape equal to $5.0$ and a rate of $10.0$, resulting in an expected relative rate of $0.5$. 
Then, using the program Seq-Gen \cite{rambaut97}, we produced DNA sequences, each 1000 bps, under the general-time reversible model for nucleotide substitution \cite{tavare86} with gamma-distributed site-rate heterogeneity \cite{yang93,yang94a}. 
The parameters of the GTR+$\Gamma$ model were drawn from the following distributions:
\begin{eqnarray*}
\boldsymbol{\eta} = (\eta_{AC},\eta_{AG},\eta_{AT},\eta_{CG},\eta_{CT},\eta_{GT}) & \sim & \mbox{Dirichlet}(2,2,2,2,2,2)\\
\boldsymbol{\pi} = (\pi_{A},\pi_{C},\pi_{G},\pi_{T}) & \sim & \mbox{Dirichlet}(10,10,10,10)\\
\alpha & \sim & \mbox{Gamma}(8,0.125),
\end{eqnarray*}
where $\boldsymbol{\eta}$ denotes the vector of relative exchangeability rates between nucleotides, $\boldsymbol{\pi}$ contains the base frequencies, and $\alpha$ is the shape parameter of the mean-one gamma distribution on site rates. 
Molecular sequence alignments were generated for each simulated tree, across our 3 different simulation conditions -- (A) $r = 0.1$, (B) $r = 0.5$, and (C) $r = 0.9$ -- resulting in 300 alignments.

\subsubsection{Random Fossil Recovery} 
Without question, fossils available for calibrating biological datasets never represent the absolute fossil history.
We addressed this for each set of simulations by randomly sampling a percentage, $\omega$, of the total fossils. 
This strategy produced four sets of calibration fossils for each simulation replicate: $\omega_5 = 5\%$, $\omega_{10} = 10\%$, $\omega_{25} = 25\%$, $\omega_{50} = 50\%$; and was applied across the three different simulation conditions (A, B, C). 

Because we planned to compare divergence-time estimates under the FBD model to calibration-density approaches, we additionally constructed a set of calibration fossils.
The calibration fossils were taken from the $\omega_{10}$ sample by selecting the oldest fossil specimen available for each calibrated node. 
Additionally, if fossil $f$ was assigned to date node $i$ and fossil $g$ was assigned to node $j$, fossil $f$ was removed if $x_i>x_j$ and $y_f < y_g$.
Calibration-density approaches condense the available information in the fossil record, thus, the 
$\omega_{cal10}$ set of fossils resulted in fewer calibrating ages compared with the $\omega_{10}$ set (Table \ref{simFossNumsTable}).  

\subsubsection{Preservation-Biased Fossil Recovery}\label{Pres-Foss-Samp-Meth}
Lineage-based variation in taphonomic properties is a well-known characteristic of the fossil record. 
We emulated this pattern by sampling fossils in proportion to a phenotypic character.
The preservation rate, $\kappa$, is a continuous-valued character, simulated  for each fossil in the absolute set of fossils according to a geometric Brownian motion model. 
For each simulation replicate, the $\kappa$ values were evolved over the complete tree by starting with $\kappa=0.5$ at the root node. 
Then, proceeding toward the tips of the tree, the value of $\kappa$ was sampled for each fossil and each internal node. 
For any given fossil or node $i$, the value $\kappa_i$ was drawn from a log-normal distribution whereby the mean was equal to the value assigned to the most recent ancestral fossil or node $\kappa_i'$. 
The variance of the log-normal distribution on $\kappa_i$ was equal to the product of the Brownian motion parameter $v$ and the difference in time between element $i$ and its nearest ancestor ($t_i$), thus $\kappa_i \sim \mbox{LN}(\ln[\kappa_i'] - \frac{\sigma^2}{2},\sigma)$, where the variance is $\sigma^2 = vt_i$. 
This continuous-trait model produces autocorrelated preservation rates over the tree, where closely related fossils share similar rates, and is identical to the autocorrelated log-normal relaxed clock model described by Thorne and Kishino \cite{thorne02} (also see \cite{thorne98,kishino01,thorne05}). 

Using the sampling/preservation rate $\kappa$, we created sets of fossils with different values of $\omega$: 5\%, 10\%, 25\%, 50\%; and applied this sampling strategy to replicates in our three different sets of simulations: (A) $r = 0.1$, (B) $r = 0.5$, and (C) $r = 0.9$. 
In contrast to the randomly sampled fossils, these sets were assembled such that fossils with larger values of $\kappa$ (relative to all other fossils for a single replicate) had a higher probability of inclusion in the subset.
Therefore, the preservation-biased sets of sampled fossils exhibited greater ascertainment bias compared with the randomly recovered fossils (for a given value of $\omega$, the random and biased sets contained the same number of fossils). 
Additionally, we created a set of calibration fossils from the $\omega_{10}$ biased subset, by retaining only the oldest fossil assigned to each calibration node.

\subsubsection{Stratigraphic Fossil Recovery}\label{Strat-Foss-Samp-Meth}
The highly partitioned nature of the geologic record is an important property of empirically sampled fossils. 
Fossil specimens are collected from discrete strata of sedimentary rock, resulting in a non-continuous historical sample. 
Sampling from distinct fossiliferous horizons violates the FBD model assumption that fossils are recovered continuously and randomly over time. 
To examine the effect of stratigraphic sampling, we created three strata for each simulation replicate where $r = 0.5$ (simulation condition B). 
For a given complete tree, the midpoints of each fossil horizon were positioned relative to the age of the root ($x_1$). 
The first stratum was closest to the root and centered at $\mathcal{S}_1 = \frac{3}{4}x_1$. 
The second interval was placed midway between the root and the tips of the tree at $\mathcal{S}_2 = \frac{1}{2}x_1$. 
The midpoint of the youngest horizon occurred near the tips of the tree at  $\mathcal{S}_3 = \frac{1}{4}x_1$. 
Then, the range of each stratigraphic layer was calculated such that it extended 10\% on either side of the midpoint, where for a given stratum with midpoint $\mathcal{S}_i$ the range was equal to $\mathcal{S}_i\pm \frac{1}{10}\mathcal{S}_i$ (Fig.~\ref{simStratigResultsFig}A).

We applied these discrete sampling horizons to generate four sets of fossils for each simulation replicate. 
The first set included all of the fossils (from the complete set) in layer $\mathcal{S}_1$, the second set comprised all fossils from $\mathcal{S}_2$, and the third set contained all fossils within $\mathcal{S}_3$. 
The fourth set of fossils was assembled by sampling 50\% of the fossils in $\mathcal{S}_1$ and 50\% of the fossils in $\mathcal{S}_3$.

\subsubsection{Non-Random Extant Species Sampling}\label{Ext-Spp-Samp-Meth}
Constant-rate birth-death processes like the SSBD or FBD models account for incomplete sampling of extant species with the parameter $\rho$. 
However, if $\rho<1$, these models assume that tip lineages are sampled uniformly \cite{yang97b,stadler09}. 
Furthermore, our implementation of the FBD model (see below) makes the extreme assumption that all extant taxa are represented with $\rho=1$. 
The assumption of random or complete taxon sampling is patently violated by most biological datasets. 
For a given group of interest, systematists might select taxa to represent the diversity of the group -- \textit{e.g.}, sampling one species per genus for a given family. 
Alternatively, researchers may apply concentrated sampling within their ingroup, but then include one or two relatively distant species to represent an outgroup, resulting in trees that are significantly imbalanced at older nodes. 
To examine the robustness of node-age estimates under the FBD model when the assumption of random or complete extant sampling is violated, we emulated the two sampling strategies described above using the simulations generated with $r=0.5$ (simulation condition B).

We subsampled the extant trees and sequence alignments to produce a phylogenetic dataset with maximized diversity by selecting tip lineages to represent the 13 oldest nodes in each replicate tree.
This was achieved by selecting the time when the 14th lineage appeared in the extant species phylogeny, then collapsing each clade descending this time point to a single lineage.
Deep-node sampling resulted in trees with 14 (out of 25) extant tips, and mimics phylogenies on higher taxa, thus violating the assumption of random species sampling \cite{hohna11}.

To automate `outgroup' sampling for a given simulation replicate, we identified the node in the extant tree with 3--12 extant descendants that also had the fewest nodes between it and the root.
From this node, we sampled one extant species. 
Thus, this sampling strategy increased the average nodal imbalance of each tree and resulted in many datasets with a densely sampled ingroup sub-tree and a lineage with just one or two species as outgroups. 
To illustrate the effect of outgroup sampling on nodal imbalance, we computed the weighted imbalance $\mathcal{U}_w$ for each node in the complete and outgroup-sampled trees. 
This tree-shape statistic takes values between 0 and 1 for any node with more than three extant descendants and
the weighted average of $\mathcal{U}_w$ across all nodes within a tree has an expectation of 0.5 under the constant-rate birth-death model \cite{fusco1995,purvis2002,heath08}. 
The average root-node imbalance in the complete extant trees was $\mathcal{U}_w=0.46$ (with a weighted average equal to 0.5). 
When extant species are removed to emulate outgroup sampling, the average root-node imbalance increased to $\mathcal{U}_w=0.97$ (with a weighted average equal to 0.6).

We used tools in the Python package DendroPy \cite{sukumaran10} to subsample extant species under these two strategies and to assign fossils to the tree. 
The fossils from the $\omega_{10}$ random subset were used to calibrate divergence times for these trees under the FBD model.
Some calibration nodes in the completely sampled trees were removed with reduced taxon sampling.
Accordingly, the fossils calibrating an absent node were assigned the the next available ancestral node when applied to the sub-sampled tree. 
Therefore, only the trees and sequence alignments were altered and the $\omega_{10}$ calibrating fossils were the same as applied to the fully sampled trees described above. 

\subsection{Simulation Study: Divergence-Time Analyses}
\subsubsection{General Priors and MCMC Details}
We estimated species divergence times for each simulation replicate and fossil subset using the Bayesian inference program DPPDiv \cite{heath11,heath12,Darriba2013}. 
Each analysis, regardless of the calibration model, conditioned divergence times on the true extant species topology, assuming a strict molecular clock with a GTR+$\Gamma$ substitution model. 
We applied standard prior distributions on the parameters and hyperparameters of the clock and substitution models \cite{heath11}. 

Each analysis was run using a single Markov chain of 2,000,000 iterations, sampling every 100th step, with the first 500,000 iterations discarded as burn-in prior to summarizing the MCMC samples. 
Because we ran approximately 2,200 MCMC analyses, it was not feasible to perform convergence diagnostics on every replicate analysis. 
However, since we evaluated performance under a unified framework, where all models are implemented in the same program with consistent priors and sampling mechanisms for overlapping parameters, summary statistics over 100 replicates are very informative about the accuracy and precision of node age estimates across our different simulation treatments and analyses.
Nevertheless, for a subset of our analyses (5 for each type of analysis), we evaluated the MCMC samples in the program Tracer \cite{rambaut09} and confirmed that the Markov chains effectively sampled the stationary distributions.

\subsubsection{Node-Age Inference under the FBD model}\label{MethCalFBDAn}
We performed divergence-time analyses under the FBD model on every simulation replicate for each fossil-sampling and taxon-sampling treatment.
Our implementation of the FBD model has four hyperparameters: $d$, $r$, $s$, and $\rho$. 
Because we are only interested in estimating node ages and not in inferring the parameters of the diversification model, we chose uniform prior densities on $d$, $r$, and $s$ to marginalize over a wide range of possible values. 
The diversification rate $d$ can take any value on the interval $(0,\infty)$, however, because they are not likely in nature, we did not simulate under extremely large values.
Thus, we place a proper, uniform prior distribution on $d$ with an arbitrarily chosen upper limit: Unif(0,30000). 
The turnover ($r$) and fossilization ($s$) parameters can each only take values on the interval (0,1) and we simply chose Unif(0,1) prior densities for each of these parameters. 
The bulk of our simulated datasets included all extant taxa, therefore we fixed the probability of sampling parameter to $\rho=1$. 
It is important to note, however, that even though we assume that the uniform priors on the diversification parameters are noninformative, in reality, such prior densities --- particularly diffuse, truncated uniform priors --- are often highly informative because they place significant prior mass on regions in parameter space with very low posterior probability \cite{felsenstein04,Yang2005}. 
Accordingly, future implementations of this model will include development of alternative hyperprior densities for these parameters. 
Nevertheless, our simulation analyses indicate that node-age estimates are robust to these uniform hyperpriors.

\subsubsection{Node-Age Inference using Calibration Densities}\label{MethCalDenAn}
Calibration density approaches require that only a single fossil is assigned to a each calibrated node. 
Thus, the set of calibration fossils $\omega_{cal10}$ constructed from the 10\% sample for both the random and preservation-biased subsets were used to estimate absolute node ages using three different node-calibration densities. 
Each of the calibration-density analyses assumed an exponential prior distribution, with the methods differing in the parameterization of the prior density. 
The exponential distribution is characterized by a single rate parameter $\epsilon$, which dictates both the mean ($\epsilon^{-1}$) and variance ($\epsilon^{-2}$) of the prior density. 
Calibration priors typically describe the time duration between the calibrated node and its fossil descendant. 
The fossil acts as a hard, minimum bound on the age of the node, thus the calibration density is offset by the age of the fossil. 

The `fixed-scaled' analysis applied a fixed-parameter exponential distribution to each calibrated node, where for any node $i$ calibrated by the fossil $f$ the rate of the prior density was scaled by the age of the fossil $y_f$. 
Under this parameterization, the rate of the exponential was $\epsilon = (\frac{1}{5}y_f)^{-1}$.
Thus, for very young fossils the expected age of the calibrated node ($\mathbb{E}[x_i] = y_f + \frac{1}{5}y_f$) would be smaller compared to the expectation for nodes calibrated by older fossils. 
By scaling the calibration density based on the fossil age, we attempted to model the arbitrary parameterization of fossil prior distributions common in divergence-time estimation analyses.

The `fixed-true' calibration prior represents an ideal, albeit unrealistic case, where the density is parameterized such that the expected age of the calibrated node is equal to its \textit{true} age. 
Under this calibration prior, for any node $i$ with true age $x^*_i$ and calibrated by fossil $f$ with age $y_f$, the rate parameter of the exponential density was fixed to $\epsilon = (x^*_i - y_f)^{-1}$. 
When applied as a zero-offset calibration prior, the fixed-true parameterization is expected to result in accurate node-age estimates since the expected age of the node is equal to the true value ($\mathbb{E}[x_i] = x^*_i$).

The hierarchical calibration density approach described by Heath \cite{heath12} uses as second-order hyperprior on the rate parameters of exponential distributions. 
This method results in robust estimates of node ages and assumes that the vector of $\epsilon$-rates for all calibrated nodes is drawn from a Dirichlet process model. 
By allowing the rates of exponential calibration densities to be random variables using MCMC, calibration-node ages are sampled from a mixture of prior distributions. 
Furthermore, this approach accounts for uncertainty in these hyperparameters and reduces the user's burden with regard to parameter specification.
For these analyses, we specified a prior-mean of three for the number of $\epsilon$-rate categories and set the remaining hyperparameters to those of Heath \cite{heath12}. 

All calibration-density methods require a prior on node ages.
This prior is applied to both calibrated and uncalibrated nodes and describes the distribution of speciation events over time. 
For all analyses using calibration priors (fixed-scaled, fixed-true, and hyperprior), we assumed a constant-rate reconstructed birth-death process \cite{gernhard08,stadler09} as a prior on speciation times. 
This model has three hyperparameters on which we applied the following priors: $d\sim\mbox{Unif}(0,30000)$, $r\sim\mbox{Unif}(0,1)$, and $\rho=1$.

\section{Acknowledgements}

We thank Paul Lewis, Mark Holder, Bastien Boussau, Michael Landis, and Brian Moore for helpful conversations and comments. TAH was supported by NSF grant DEB-1256993; NIH grants GM-069801 and GM-086887 awarded to JPH.
TS thanks the Swiss National Science foundation for funding (SNF grant \#PZ00P3 136820).


\printbibliography[heading=siref]

\newpage
\section{Tables}
\setlength{\tabcolsep}{12pt}

\begin{table}[tbh]
\centering
\caption{Coverage probabilities across all nodes for different calibration approaches.}\label{summaryTableCompare}
\begin{tabular}{@{\extracolsep{\fill}}l  c c c }
\hline
  &\multicolumn{3}{@{}c}{\textbf{Coverage probability}} \\\cline{2-4}
\multicolumn{1}{@{}l}{\textbf{Calibration method}}  & \multicolumn{1}{c}{\textbf{$r=0.1$}} & \multicolumn{1}{c}{\textbf{$r=0.5$}} & \multicolumn{1}{c}{\textbf{$r=0.9$}}  \\ 
\hline
FBD ($\omega_{10}$) & 0.956 & 0.966 & 0.960\\
FBD ($\omega_{cal10}$) & 0.950 & 0.953 & 0.958 \\
Hyperprior ($\omega_{cal10}$) & 0.938 & 0.930 & 0.929 \\
Fixed--true ($\omega_{cal10}$) & 0.927 & 0.904 &  0.921 \\
Fixed--scaled ($\omega_{cal10}$) & 0.733 & 0.680 & 0.726 \\
\hline
\end{tabular}
\end{table}

\bigskip
\begin{table}[tbh]
\centering
\caption{Sequence data and GenBank accession numbers for analysis of all bears (Ursidae) and two outgroups (\textit{Canis lupus} and \textit{Phoca largha}).}\label{bearsDNAGBTable}
\begin{tabular}{@{\extracolsep{\fill}}l c c c }
\hline
& &\multicolumn{2}{c}{\textbf{Accession number}} \\ \cline{3-4}
\multicolumn{1}{@{}l}{\textbf{Species}} & \multicolumn{1}{c}{\textbf{Common name}}  & \multicolumn{1}{c}{\textbf{irbp}} & \multicolumn{1}{c}{\textbf{mtDNA}}  \\ 
\hline
\textit{Canis lupus} & gray wolf & AB499823 & AY525044 \\
\textit{Phoca largha} & spotted seal & NC\_008430 & AB188519 \\
\textit{Ailuropoda melanoleuca} & giant panda & NC\_009492 & AY303836 \\
\textit{Tremarctos ornatus} & spectacled bear & NC\_009969 & AY303840 \\
\textit{Melursus ursinus} & sloth bear & NC\_009970 & AY303838 \\
\textit{Helarctos malayanus} & sun bear & NC\_009968 & AY303839 \\
\textit{Ursus arctos} & brown bear & NC\_003427 & AY303842 \\
\textit{Ursus maritimus} & polar bear & NC\_003428 & AY303843 \\
\textit{Ursus americanus} & American black bear & NC\_003426 & AY303837 \\
\textit{Ursus thibetanus} & Asian black bear & NC\_009971 & AY303841 \\
\hline
\end{tabular}
\end{table}

\begin{table}[tbh!]
\centering
\caption{Fossil species used for calibrating divergence times under the FBD model.}\label{bearFossilTable}
\begin{tabular}{@{\extracolsep{\fill}}l  c c c c}
\hline
 &\multicolumn{2}{@{}c}{\textbf{Age (Ma)}} \\ \cline{2-3}
\multicolumn{1}{@{}l}{\textbf{Fossil species}}  & \multicolumn{1}{c}{\textbf{range}} & \multicolumn{1}{c}{\textbf{random$^{*}$}} & \multicolumn{1}{c}{\textbf{Node$^{**}$}} & \multicolumn{1}{c}{\textbf{Citation}} \\ 
\hline
Canidae & \\
\hspace{2mm} \textit{Hesperocyon gregarius} & 37.2--40 & 39.07 & 1 & \cite{wang1994,wang1999}\\
\hspace{2mm} \textit{Caedocyon tedfordi} & 20.43--30.8 & 25.88 & 1
 &  \cite{wang1994,wang1999}\\
\hspace{2mm} \textit{Osbornodon sesnoni} & 20.43--30.8 & 30.08 & 1 &  \cite{wang1994,wang1999}\\
\hspace{2mm} \textit{Cormocyon copei} & 24.8--26.3 & 26.14 & 1 &  \cite{wang1994,wang1999}\\
\hspace{2mm} \textit{Borophagus diversidens} & 1.8--4.9 & 4.248 & 1 &  \cite{wang1994,wang1999}\\
& & \\
Pinnipedimorpha & \\
\hspace{2mm} \textit{Enaliarctos tedfordi} & 25--28 & 27.11 & 2 & \cite{fulton2010}\\
\hspace{2mm} \textit{Proneotherium repenningi} & 15.97--20 & 17.92 & 2 & \cite{fulton2010}\\
\hspace{2mm} \textit{Leptophoca lenis} & 14.2--16.3 & 14.99 & 2 & \cite{fulton2010}\\
\hspace{2mm} \textit{Acrophoca} & 5--7 & 6.695 & 2 & \cite{fulton2010}\\
\hspace{2mm} \textit{Phoca vitulina} & 0.79--1.64 & 0.805 & 2 & \cite{fulton2010}\\
& & \\
Ursidae & \\
\hspace{2mm} \textit{Parictis montanus} & 33.9--37.2 & 36.60 & 2 & \cite{clark1972,krause2008}\\
\hspace{2mm} \textit{Zaragocyon daamsi} & 20--22.8 & 21.86 & 2 & \cite{ginsburg1995,abella12}\\
\hspace{2mm} \textit{Ballusia elmensis} & 13.7--16 & 14.01 & 2 & \cite{ginsburg1998,abella12}\\
\hspace{2mm} \textit{Ursavus primaevus} & 13.65--15.97 & 14.41 & 2 & \cite{andrews1977,abella12}\\
\hspace{2mm} \textit{Ursavus brevihinus} & 15.97--16.9 & 16.20 & 2 & \cite{heizmann1980,abella12}\\
\hspace{2mm} \textit{Indarctos vireti} & 7.75--8.7 & 8.680 & 3 & \cite{montoya2001,abella12}\\
\hspace{2mm} \textit{Indarctos arctoides} & 8.7--9.7 & 9.545 & 3 & \cite{geraads2005,abella12}\\
\hspace{2mm} \textit{Indarctos punjabiensis} & 4.9--9.7 & 4.996 & 3 & \cite{baryshnikov2002,abella12}\\
\hspace{2mm} \textit{Ailurarctos lufengensis} & 5.8--8.2 & 7.652 & 3 & \cite{jin2007,abella12}\\
\hspace{2mm} \textit{Agriarctos spp.} & 4.9--7.75 & 5.006 & 3 & \cite{abella2011,abella12}\\
\hspace{2mm} \textit{Kretzoiarctos beatrix} & 11.2--11.8 & 11.69 & 3 & \cite{abella2011,abella12}\\
\hspace{2mm} \textit{Arctodus simus} & 0.012--2.588 & 0.487 & 5 & \cite{churcher1993,krause2008}\\
\hspace{2mm} \textit{Ursus abstrusus} & 1.8--5.3 & 4.27 & 6 & \cite{bjork1970,krause2008}\\
\hspace{2mm} \textit{Ursus spelaeus} & 0.027--0.25 & 0.054 & 6 & \cite{loreille2001,krause2008}\\
\hline
\multicolumn{5}{l}{%
  \begin{minipage}{14cm}%
    \footnotesize $^{*}$Since most fossil species are associated with age ranges, a random age (sampled from a uniform distribution on the age range) was used for the calibration age for each fossil. This random age is intended to approximate random fossil sampling since the FBD model assumes that the fossil represents a single point in time. \\$^{**}$The node corresponding to the labels in Fig.~\ref{bearCompareDivtimeFig} that is the ancestor of the fossil.%
  \end{minipage}%
}\\
\end{tabular}
\end{table}
\clearpage
\newpage

\begin{table}[tbh]
\centering
\caption{Mean and range of the numbers of simulated fossils.}\label{simFossNumsTable}
\begin{tabular}{@{\extracolsep{\fill}}l r r r }
\hline
 &\multicolumn{3}{@{}c}{\textbf{Turnover rate}} \\ \cline{2-4}
\multicolumn{1}{@{}l}{\textbf{Fossil sample}} & \multicolumn{1}{c}{\textbf{(A) $r = 0.1$}}  & \multicolumn{1}{c}{\textbf{(B) $r = 0.5$}} & \multicolumn{1}{c}{\textbf{(C) $r = 0.9$}}   \\ 
\hline
$100\%$ & 176.87 [79, 294] & 190.47 [101, 462] & 272.88 [59, 1991] \\
$\omega_{50}$ & 88.72 [40, 147] & 95.45 [50, 231] & 136.69 [30, 996]\\
$\omega_{25}$ & 44.31 [20, 74] & 47.79 [25, 116] & 68.35 [15, 498] \\
$\omega_{10}$ & 17.74 [8, 29] & 19.08 [10, 46] & 27.34 [6, 199]\\
$\omega_{cal10}$ & 9.51 [4, 14] & 9.23 [5, 14]  & 7.53 [4, 12] \\
$\omega_{5}$ & 8.86 [4, 15] & 9.57 [5, 23]  & 13.65 [3, 100] \\
\hline
\multicolumn{4}{l}{%
  \begin{minipage}{12.5cm}%
    \footnotesize The turnover rate is $r = \mu/\lambda$.%
  \end{minipage}%
}\\
\end{tabular}
\end{table}

\bigskip
\begin{table}[tbh]
\centering
\caption{Notation used to state FBD tree probability.}\label{FBDMathNotation}
\begin{tabular}{@{\extracolsep{\fill}}l l }
\hline
\multicolumn{1}{@{}l}{\textbf{Parameter}} &  \multicolumn{1}{c}{\textbf{Definition}}  \\ 
\hline
$n$ & Number of extant species \\
$m$ & Number of fossil samples \\
$k$ & Number of $m$ fossils that occur on branches in the extant tree  \\
$m-k$ & Number of $m$ fossils attach to the sampled tree via speciation  \\
$\mathcal{T}$ & FBD tree topology and branching times for $n$ species and $m$ fossils \\
$\mathcal{V} = (1,\ldots,n-1)$ & Vector of internal node indices, labeled in preorder sequence \\
$x_i$ & Age of internal node $i$ (for $i\in\mathcal{V}$)\\
$\mathcal{F} = (1,\ldots,m)$ & Vector of calibrating fossils \\
$y_f$ & Age of fossil specimen $f$ (for $f\in\mathcal{F}$)\\
$z_f$ & Attachment time of fossil specimen $f$ \\
$\gamma_f$ & number of possible attachment lineages of fossil specimen $f$ \\
$\mathcal{I}_f$ & Indicator function for fossil $f$, ${\cal I} = 0$ if fossil is ancestral, else ${\cal I} = 1$ \\
$\lambda$ & Speciation rate \\
$\mu$ & Extinction rate \\
$\psi$ & Fossil recovery rate \\
$\rho$ & Extant species sampling probability \\
\hline
\end{tabular} 
\end{table}

\newpage
\FloatBarrier
\section{Figures}

\end{document}